\documentclass[a4paper,11pt]{article}
\pdfoutput=1 % if your are submitting a pdflatex (i.e. if you have

\usepackage{jheppub}

\usepackage{graphicx}
\usepackage{amssymb}
\usepackage{amsmath}

\usepackage[dvipsnames]{xcolor}
\usepackage{hyperref}

\definecolor{darkblue}{RGB}{0, 70, 150}
\definecolor{darkred}{RGB}{200, 0, 0}
\definecolor{darkgreen}{RGB}{0, 180, 0}
\definecolor{darkyellow}{RGB}{210, 210, 0}

\newcommand{\hc}[1]{#1^{\dagger}}

\newcommand{\abs}[1]{|#1|}

\newcommand{\diag}{\operatorname{diag}}

\newcommand{\MM}{\left|\bar M \right|^2}

\newcommand{\diracd}[4]{\delta^{(4)}(#1\!+\!#2\!-\!#3\!-\!#4)}

\title{Phase Transitions and Gravitational Waves in a Model of $\mathbb{Z}_{3}$ Scalar Dark Matter}

\author[a]{Nico Benincasa,}
\author[b]{Andrzej Hryczuk,}
\author[a]{Kristjan Kannike,}
\author[c]{Maxim Laletin}

\affiliation[a]{National Institute of Chemical Physics and Biophysics, \\
R\"{a}vala 10, Tallinn, Estonia}
\affiliation[b]{National Centre for Nuclear Research, Pasteura 7, 02-093 Warsaw, Poland}
\affiliation[c]{Institute of Theoretical Physics, Faculty of Physics, University of Warsaw, Pasteura 5, 02-093 Warsaw, Poland}

\emailAdd{nico.benincasa@kbfi.ee}
\emailAdd{andrzej.hryczuk@ncbj.gov.pl}
\emailAdd{Kristjan.Kannike@cern.ch}
\emailAdd{maxim.laletin@fuw.edu.pl}

\date{\today}

\abstract{
 Theories with more than one scalar field often exhibit phase transitions producing potentially detectable gravitational wave (GW) signal. In this work we study the semi-annihilating $\mathbb{Z}_3$ dark matter model, whose dark sector comprises an inert doublet and a complex singlet, and assess its prospects in future GW detectors. Without imposing limits from requirement of providing a viable dark matter candidate, i.e. taking into account only other experimental and theoretical constraints,
 we find that the first order phase transition in this model can be strong enough to lead to a detectable signal. However, direct detection and the dark matter thermal relic density constraint calculated with the state-of-the-art method including the impact of early kinetic decoupling, very strongly limit the parameter space of the model explaining all of dark matter \textit{and} providing observable GW peak amplitude. Extending the analysis to underabundant dark matter thus reveals region with detectable GWs from a single-step or multi-step phase transition.
}

\begin{document}
\maketitle

%%%%%%%%%%%%%%%%%%%%%%%%%%%%%%%%%%%%%%%%%%%%%%%%%%%%%%%%%%
\section{Introduction}

Although dark matter (DM) constitutes 27\% of the energy density of the Universe \cite{Planck:2018vyg}, its nature is still unknown. 
The discovery of the Higgs boson, over a decade ago, showed that scalars do exist in nature \cite{CMS:2012qbp,ATLAS:2012yve}. In particular, there are two well-motivated candidates of scalar dark matter: the inert doublet  \cite{Ma:2006km,LopezHonorez:2006gr,Deshpande:1977rw,Barbieri:2006dq} and the scalar singlet \cite{Silveira:1985rk,McDonald:1993ex,Burgess:2000yq}, usually stabilised by a $\mathbb{Z}_{2}$ parity. Both these DM candidates suffer, however, in the face of the most recent results from the direct detection experiments XENON1T \cite{XENON:2018voc}, PandaX-4T \cite{PandaX-4T:2021bab} and LZ \cite{LZ:2022ufs} that have this far not seen any evidence of a dark-matter signal.

A possible solution lies in considering a larger scalar dark sector comprising both the inert doublet and a complex singlet with a stabilising symmetry more complex than the $\mathbb{Z}_{2}$ parity. In this case, dominant processes that determine the DM relic density can belong to semi-annihilation \cite{Hambye:2009fg,Hambye:2008bq,DEramo:2012fou,DEramo:2010ep} instead of the usual annihilation. Since semi-annihilation and direct-detection processes are driven by different couplings, the correct DM relic density may be obtained while the spin-independent direct-detection cross section lies below the current limits of direct-detection experiments. 

In fact, semi-annihilation is possible already for the next-to-simplest stabilising symmetry, the  $\mathbb{Z}_{3}$ group. In the $\mathbb{Z}_3$ complex singlet-inert doublet model, the impact of semi-annihilations on DM relic density  was first considered in \cite{Belanger:2012vp} and a comprehensive study of the model was performed in \cite{Belanger:2014bga}.\footnote{The $\mathbb{Z}_{2}$ model with the same field content was studied in \cite{Kadastik:2009gx,Kadastik:2009dj,Kadastik:2009cu,Kadastik:2009ca}. More complicated symmetries may further restrict possible interactions: see e.g. \cite{Aoki:2014cja} and \cite{Kakizaki:2016dza}. With a $\mathbb{Z}_{4}$ symmetry one can have two-component dark matter for the same field content \cite{Belanger:2012vp,Belanger:2014bga,Belanger:2021lwd,Belanger:2022qxt} (see also e.g. \cite{Yaguna:2019cvp,BasiBeneito:2020mdr,Yaguna:2021vhb,Jurciukonis:2022oru} for generalisations). 
With higher symmetries, such as $\mathbb{Z}_7$, and more fields, one can have even three-component DM \cite{Belanger:2022esk}.}
 On the other hand, the $\mathbb{Z}_3$ singlet model \cite{Belanger:2012zr,Bhattacharya:2017fid}\footnote{It is possible that the $\mathbb{Z}_{3}$ symmetry is a remnant of breaking of a $U(1)$ group, see e.g. \cite{Ko:2014nha,Ko:2014loa,Choi:2015bya,Ma:2015mjd}. On the other hand, if the $\mathbb{Z}_3$ symmetry itself is broken, but the CP symmetry is conserved, the pseudoscalar part of the complex singlet is still stable and can be a DM candidate \cite{Kannike:2019mzk,Cheng:2022hcm}.} and the inert doublet model \cite{Ma:2006km,LopezHonorez:2006gr,Deshpande:1977rw,Barbieri:2006dq} can be considered as limiting cases of the full model. The $\mathbb{Z}_3$ singlet model was reconsidered with improved bounds and impact of early kinetic decoupling in \cite{Hektor:2019ote}. Indirect detection signals have been considered in \cite{Cai:2015tam,Arcadi:2017vis,Queiroz:2019acr} and  cosmic phase transitions were considered in \cite{Kang:2017mkl}. A fit of the $\mathbb{Z}_3$ singlet model was made in \cite{Athron:2018ipf}. The $\mathbb{Z}_{3}$ singlet-doublet model also shares many constraints with the inert doublet model (IDM) \cite{Swiezewska:2012eh,Krawczyk:2013jta,Belyaev:2016lok,Belyaev:2018ext}. Recent studies of DM phenomenology in the IDM are given by \cite{Goudelis:2013uca,Diaz:2015pyv,Blinov:2015qva}. The phase structure of the IDM was studied in  \cite{Ginzburg:2010wa}. First-order phase transitions in the IDM were studied in \cite{Gil:2012ya,Borah:2012pu,Cline:2013bln,Blinov:2015vma,Fabian:2020hny} and the GW phenomenology of the inert doublet model was elucidated in \cite{Benincasa:2022elt}. IDM, augmented by higher-order operators, can also given rise to baryogenesis \cite{Astros:2023gda}. Notably, $\mathbb{Z}_3$ symmetry and semi-annihilation have also been considered for a singlet-extended type II seesaw model \cite{Ghosh:2022fzp}.

In the part of the parameter space where semi-annihilation dominates, one has to take extra care in calculating the DM relic density. In the standard calculation of the thermal relic abundance \cite{Gondolo:1990dk} it is assumed that, at the freeze-out, DM is still in equilibrium with the heat bath. Indeed, usually the elastic scattering processes with the thermal bath particles are much more efficient than the processes of annihilation and production. If the latter are enhanced, however, or if the scattering processes are not related to the number changing ones -- which is the case for semi-annihilation -- this assumption may not be satisfied. It has been shown that kinetic decoupling can begin as early as the chemical one and that the change in the DM phase space distribution may modify the relic abundance by more than an order of magnitude \cite{Binder:2017rgn,Duch:2017nbe,Binder:2021bmg}.

Although direct detection of DM remains elusive, other observational channels are becoming available. The discovery of gravitational waves (GW) by the LIGO experiment \cite{LIGOScientific:2016sjg,LIGOScientific:2016aoc} has begun a new era of GW astronomy. Of particular interest to particle physicists is the possibility to probe the vacuum structure of the scalar potential via cosmic phase transitions. In the SM, the phase transition is a smooth cross-over \cite{Kajantie:1996mn,Aoki:1999fi} which does not produce a GW signal. A larger scalar sector in an extension of the SM, however, can give rise to first-order phase transitions \cite{Witten:1984rs,Steinhardt:1981ct,Hogan:1983ixn} which can produce a stochastic GW background that could be detected by future GW detectors such as LISA \cite{LISA:2017pwj,eLISA:2013xep}, BBO \cite{Crowder:2005nr,Corbin:2005ny}, Taiji \cite{Ruan:2018tsw,Hu:2017mde}, TianQin \cite{TianQin:2015yph} or DECIGO \cite{Seto:2001qf,Kawamura:2020pcg} (see e.g. Ref. \cite{Athron:2023xlk} for a pedagogical review).

The aims of this paper are therefore twofold: first of all to study gravitational-wave signals originating during phase transitions predicted by the  inert doublet-singlet model with a $\mathbb{Z}_{3}$ symmetry and, second, to update the phenomenological analysis of this model by including constraints from vacuum stability, unitarity and perturbativity, and electroweak precision parameters and calculating the thermal relic density taking into account the effects of early kinetic decoupling.

The paper is organised as follows. We describe the tree-level scalar potential of the model together with the loop and thermal corrections in Section~\ref{sec:model}. Various theoretical and experimental constraints are given in Section~\ref{sec:constraints}. Relic density calculation with early kinetic decoupling is discussed in Section~\ref{sec:relic}. Calculations of phase transitions and the gravitational-wave signal are elucidated in Section~\ref{sec:pt:gw}. Our Monte Carlo scan is described in Section~\ref{sec:scan} and the signals in Section~\ref{sec:signals}. We conclude in Section~\ref{sec:conclusions}. Appendices list field-dependent scalar masses (\ref{sec:mass_matrix}), effective potential counter-terms (\ref{sec:counterterms}) and formulae for the sound-wave contribution to the gravitational-wave signal (\ref{sec:sw_contribution}).

%%%%%%%%%%%%%%%%%%%%%%%%%%%%%%%%%%%%%%%%%%%%%%%%%%%%%%%%%%
\section{$\mathbb{Z}_{3}$ dark matter model}
\label{sec:model}

%%%%%%%%%%%%%%%%%%%%%%%%%%%%%%%%%%%%%%%%%%%%%%%%%%%%%%%%%%
\subsection{Scalar potential and parametrisation}
\label{sec:pot}

Our scalar fields comprise, besides the Standard-Model Higgs boson $H_{1}$, an inert doublet $H_{2}$ and a complex singlet $S$. Dark matter is made stable thanks to a $\mathbb{Z}_{3}$ discrete symmetry under which the Higgs boson, together with all other SM fields, transforms trivially: $H_{1} \to H_{1}$, $S \to \omega S$, $H_{2} \to \omega H_{2}$ with $\omega^{3} = 1$ (considering $S \to \omega^{2} S$, $H_{2} \to \omega^{2} H_{2}$ is equivalent). The most general scalar potential invariant under the $\mathbb{Z}_{3}$ is given by \cite{Belanger:2012vp,Belanger:2014bga}
\begin{equation}
\begin{split}
  V &=  \mu_{1}^{2} \abs{H_{1}}^{2} + \lambda_{1} \abs{H_{1}}^{4} 
  + \mu_{2}^{2} |H_{2}|^{2} + \lambda_{2} |H_{2}|^{4} + \mu_{S}^{2} |S|^{2} + \lambda_{S} |S|^{4} \\
  &+ \lambda_{S1} |S|^{2} |H_{1}|^{2}
  + \lambda_{S2} |S|^{2} |H_{2}|^{2} + \lambda_{3} |H_{1}|^{2} |H_{2}|^{2} 
  + \lambda_{4} (H_{1}^{\dagger} H_{2}) (H_{2}^{\dagger} H_{1})
  \\
  &+ \frac{\mu''_{S}}{2} (S^{3} + S^{\dagger 3}) 
  + \frac{\lambda_{S12}}{2} (S^{2} H_{1}^{\dagger}   H_{2} + S^{\dagger 2} H_{2}^{\dagger} H_{1}) + \frac{\mu_{SH}}{2} (S H_{2}^{\dagger} H_{1} + S^{\dagger} H_{1}^{\dagger} H_{2}),
\end{split}
\label{eq:V:c}
\end{equation}
where we take all the parameters real.
We can parametrise the fields in the EW vacuum in the Landau gauge as
\begin{equation}
  H_1 = 
  \begin{pmatrix}
  G^{+}
  \\
  \frac{v + h + i G^{0}}{\sqrt{2}}
  \end{pmatrix}, 
  \quad 
  H_2 = 
  \begin{pmatrix}
  H^{+} 
  \\
  \frac{H + i A}{\sqrt{2}}
  \end{pmatrix}, 
  \quad 
  S = \frac{s + i\chi}{\sqrt{2}},
\end{equation}
where $v = 246.22~\text{GeV}$ is the vacuum expectation value (VEV) of the SM Higgs field $H_{1}$ at zero temperature.

Notice that the absence of the term $\lambda_{5}[(H_{1}^{\dagger} H_{2})^2+(H_{2}^{\dagger} H_{1})^2]$
%$\lambda_{5}\left[\left(H_{1}^{\dagger} H_{2}\right)^2+\left(H_{2}^{\dagger} H_{1}\right)^2\right]$
-- forbidden by the $\mathbb{Z}_{3}$ symmetry -- means that there is no mass splitting between the scalar component $H$ and pseudoscalar component $A$ of the neutral part of $H_{2}$.
The $\mu_{SH}$ term in the above potential induces a mixing by an angle $\theta$ between $S$ and the neutral part of $H_2$. In terms of the complex mass eigenstates $x_1$ and $x_2$, it yields
\begin{equation}
\frac{H + iA}{\sqrt{2}} = x_1\sin\theta +x_2\cos\theta\quad\text{and}\quad \frac{s + i\chi}{\sqrt{2}} = x_1\cos\theta -x_2\sin\theta.
\end{equation}
Because $x_{2}$ is doublet-like and has gauge interactions with the $Z$ boson, it cannot be a DM candidate due to its large direct detection cross section. Therefore, we always consider $M_{x_{1}} < M_{x_{2}}$ and choose the singlet-like $x_{1}$ as our DM candidate.

We then express the parameters in Eq.~(\ref{eq:V:c}) in terms of the physical masses $M_{h}$, $M_{x_1}$, $M_{x_2}$, $M_{H^\pm}$, the dark sector mixing angle $\theta$ and the Higgs VEV $v$:
\begin{align}
    \mu_{1}^2 &= -\frac{M_{h}^2}{2}, \quad\mu_2^2 = M_{H^\pm}^2-\frac{\lambda_3 v^2}{2}, \quad\mu_S^2 = M_{x_1}^2 \cos^2\theta + M_{x_2}^2 \sin^2\theta -\frac{\lambda_{S1}v^2}{2},    \nonumber \\
    \mu_{SH} &= \frac{\sqrt{2}}{v}(M_{x_1}^2-M_{x_2}^2)\sin 2\theta,
    \quad \lambda_1 = \frac{M_{h}^2}{2v^2} ,\quad \lambda_4 = \frac{2}{v^2} (M_{x_1}^2 \sin^2\theta + M_{x_2}^2 \cos^2\theta - M_{H^\pm}^2).
    \label{eq:parametrization}
\end{align}

%%%%%%%%%%%%%%%%%%%%%%%%%%%%%%%%%%%%%%%%%%%%%%%%%%%%%%%%%%
\subsection{Potential for phase transitions}
\label{sec:qc}

%%%%%%%%%%%%%%%%%%%%%%%%%%%%%%%%%%%%%%%%%%%%%%%%%%%%%%%%%%
\subsubsection{Tree-level potential for phase transitions}

When dealing with phase transitions and gravitational-wave signals in Subsection~\ref{sec:GW}, we consider the path to tunnel from the origin to the EW minimum to lie in the three-dimensional field space of $h, H$ and $s$. Therefore in the potential~\eqref{eq:V:c} we set all other scalar fields to zero. The resulting tree-level potential is thus
\begin{align}
V_\text{cl}(h,H,s) &= \frac{\mu_1^2}{2} h^2 +\frac{\lambda_1}{4} h^4+ \frac{\mu_2^2} {2}H^2 +\frac{\lambda_2}{4} H^4+ \frac{\mu_S^2}{2} s^2 +\frac{\lambda_S}{4} s^4+ \frac{\lambda_3+\lambda_4}{4}h^2H^2 \nonumber \\
&+\frac{\lambda_{S1}}{4} h^2s^2 +\frac{\lambda_{S2}}{4} H^2s^2 + \frac{\lambda_{S12}}{4} h H s^2 + \frac{\mu''_S}{2\sqrt{2}} s^3 +\frac{\mu_{SH}}{2\sqrt{2}}h H s.
\label{eq:Vcl}
\end{align}

%%%%%%%%%%%%%%%%%%%%%%%%%%%%%%%%%%%%%%%%%%%%%%%%%%%%%%%%%%
\subsubsection{Quantum corrections to the potential}
\label{sec:qc}

The Coleman-Weinberg potential contains the quantum corrections to the tree-level potential \eqref{eq:Vcl} given by the sum of the one-particle-irreducible diagrams with external lines from amongst the classical background fields $h, H$ and $s$. The one-loop Coleman-Weinberg potential  is expressed in the Landau gauge as~\cite{PhysRevD.7.1888}
\begin{equation}
    \label{eq:Vcw}
    \Delta V(h, H, s) = \frac{1}{64 \pi^2} \sum_i n_i m_i^4 \left( \ln \frac{m_i^2}{\mu^2} - C_i \right),
\end{equation}
where the sum ranges over the fields $i = W_T^\pm, W_L^\pm, Z_T, Z_L, \gamma_L, t, h, H, s, G_0, A, \chi, G^\pm, H^\pm$. We neglect the lighter fermions because their Yukawa couplings are very small as compared to the top Yukawa. We choose the renormalisation scale to be $\mu = v$, and $C_i$ are constants peculiar to the renormalisation scheme. The subscripts $T$ and $L$, respectively, denote the transverse and longitudinal components of the gauge bosons. The bosonic and fermionic degrees of freedom\footnote{Strictly speaking, talking about degrees of freedom $n_W, n_Z$ for the massive gauge bosons is an abuse of notation; the fortuitous correspondence between their value and the number of polarisation states is not true in every gauge~\cite{Delaunay:2007wb}.} are given
by $n_{W_T^\pm} = 4, n_{W_L^\pm} = 2, n_{Z_T} = 2, n_{Z_L} = 1,n_{\gamma_L} = 1, n_t = -12, n_h = n_H = n_s = n_{G_0} = n_A =n_{\chi}= 1$ and $n_{H^\pm} = n_{G^\pm} = 2$. The constants $C_i = 3/2$ for scalars, fermions and longitudinal vector bosons, as well as $C_i = 1/2$ for transverse vector bosons in the $\overline{\text{MS}}$ subtraction scheme. The field-dependent scalar masses $m_i^2\equiv m_i^2(h,H,s)$ correspond to the eigenvalues of the field-dependent mass matrices of the scalar fields $M^2_S$, of the pseudoscalar fields $M^2_P$ and of the charged fields $M^2_C$, which are given in the Appendix~\ref{sec:mass_matrix}. The field-dependent top quark and gauge boson masses are given by
\begin{equation}
m^2_t = \frac{y_t^2}{2}h^2,\quad m^2_{W_T^\pm}=m^2_{W_L^\pm}=\frac{g_2^2}{4}(h^2+H^2),\quad m^2_{Z_T}=m^2_{Z_L}=\frac{g_1^2+g_2^2}{4}(h^2+H^2),\quad m^2_{\gamma_L}=0,
\end{equation}
where $y_t$, $g_1$ and $g_2$ are the top Yukawa coupling, the $U(1)_Y$ and $SU(2)_L$ gauge couplings, respectively.

In order to fix the masses, VEV and the dark sector mixing angle at their tree-level values, we cancel  contributions from the one-loop $\Delta V$ in the EW vacuum with opposite-sign contributions from the  counterterm potential given by
\begin{equation}
\begin{split}
\delta V(h,H,s) &= \delta\mu_1^2 h^2 + \delta\mu_2^2 H^2 + \delta\mu_S^2 s^2 +\delta\lambda_1 h^4 + (\delta\lambda_3+\delta\lambda_4)h^2H^2 \\
&+\delta\lambda_{S1} h^2s^2 + \delta\lambda_{S12} h H s^2 + \delta\mu''_S s^3 +\delta\mu_{SH}h H s,
\label{eq:Vct}
\end{split}
\end{equation}
whose coefficients are given in the Appendix~\ref{sec:counterterms}.
This cancellation is guaranteed by the following set of renormalisation conditions: 
\begin{align}
 \partial_h \left(\Delta V+\delta V\right)\Big\vert_{\textsc{vev}} &=0, &
 \partial^2_{h^2} \left(\Delta V+\delta V\right)\Big\vert_{\textsc{vev}} &=0, &
 \partial^2_{H^2} \left(\Delta V+\delta V\right)\Big\vert_{\textsc{vev}} &=0, 
 \notag
 \\
 \partial^2_{s^2} \left(\Delta V+\delta V\right)\Big\vert_{\textsc{vev}} &=0, &
 \partial_H \partial_s \left(\Delta V+\delta V\right)\Big\vert_{\textsc{vev}} & = 0, &
 \partial_h \partial^2_{H^2} \left(\Delta V+\delta V\right)\Big\vert_{\textsc{vev}} &=0, 
 \notag 
 \\ 
 \partial_h \partial^2_{s^2} \left(\Delta V+\delta V\right)\Big\vert_{\textsc{vev}} &=0, &
 \partial_H \partial^2_{s^2} \left(\Delta V+\delta V\right)\Big\vert_{\textsc{vev}} &=0, &
 \partial^3_{s^3} \left(\Delta V+\delta V\right)\Big\vert_{\textsc{vev}} &= 0,
\label{eq:renorm_condition}
\end{align}
where \textsc{vev} means at the EW minimum $(h,H,s)=(v,0,0)$. In Eq.~(\ref{eq:renorm_condition}), we take care of the problematic IR divergence occurring in $\ln m^2_G$ (due to the vanishing Goldstone mass in the EW vacuum) by replacing $m_G$ with $M_{h}$~\cite{Cline:2011mm} as an infrared regulator.

%%%%%%%%%%%%%%%%%%%%%%%%%%%%%%%%%%%%%%%%%%%%%%%%%%%%%%%%%%
\subsubsection{Thermal corrections to the potential}

The one-loop finite-temperature contribution to the effective potential is given by~\cite{PhysRevD.9.3320}
\begin{equation}
    \label{eq:VT}
    V_{\rm T} (h, H, s, T) = \frac{T^4}{2\pi} 
    \sum_i n_i J_{\text{B/F}}\left(\frac{m_i^2}{T^2}\right),
\end{equation}
where the thermal bosonic/fermionic functions $J_{\text{B/F}}$ are defined as~\cite{PhysRevD.9.3320}
\begin{equation}
J_{\text{B/F}}(y^2) = \int_0^\infty dx~x^2 \ln\left(1\mp e^{-\sqrt{x^2+y^2}}\right).
\end{equation}
It is customary to deal with IR divergences arising from the zero bosonic Matsubara mode in the high-temperature limit by considering the so-called daisy resummation~\cite{RevModPhys.53.43}. This procedure amounts to adding the leading-order thermal correction or Debye mass $\Pi_i T^2$ to the mass $m_i^2$ in $\Delta V$ and $V_T$~\cite{Parwani:1991gq}. For the scalar sector it means
\begin{equation}
\mu_i^2 \rightarrow \mu_i^2 + c_i T^2 \text{\quad with \quad} i=1,2,S,
\end{equation} 
where
\begin{align}
c_1 &= \frac{1}{16}(g_1^2 + 3 g_2^2 + 4 y_t^2) + \frac{1}{12} (6\lambda_1 + 2\lambda_3 + \lambda_4 + \lambda_{S1}), 
\\ 
c_2 &= \frac{1}{16}(g_1^2 + 3 g_2^2) + \frac{1}{12} (6\lambda_2 + 2\lambda_3 + \lambda_4 + \lambda_{S2}),
\\  
c_S &= \frac{1}{6}(2\lambda_S + \lambda_{S1} + \lambda_{S2})
\end{align}
with the SM contribution taken from~\cite{PhysRevD.45.2933}. There is no need to consider the thermal mass of the top quark and fermions in general, because daisy diagrams do not diverge in the IR,  since there is no zero Matsubara frequency for fermions~\cite{Cline:1996mga, Braaten:1991gm, Enqvist:1997ff}. As for gauge bosons, only their longitudinal modes acquire a significant thermal correction to their mass, thereby the Debye mass of transverse modes (protected by gauge symmetry) is neglected~\cite{Cline:1996mga, Espinosa:1992kf, Manuel:1998vg}.

Notice that thermal contributions to the longitudinal part of $Z$ and $\gamma$ should be added to their mass matrix in the gauge basis before it is diagonalised. The Debye mass of $W_L$ is $\Pi_{W_L} T^2 = 2 g_2^2 T^2$, while for $Z_L$ and $\gamma_L$, one should add 
\begin{equation}
\diag(c_Z T^2, c_\gamma T^2), \quad\text{with}\quad c_Z = 2g_2^2,\quad c_\gamma = 2g_1^2
\end{equation}
to their mass matrix and then diagonalise it~\cite{Cline:1996mga, Bernon:2017jgv}.

%%%%%%%%%%%%%%%%%%%%%%%%%%%%%%%%%%%%%%%%%%%%%%%%%%%%%%%%%%
\subsubsection{Full thermally-corrected effective potential }

The full thermally-corrected effective potential  needed  to compute phase-transition parameters is thus made of the classical potential ~(\ref{eq:Vcl}), to which one adds quantum corrections summarised in~(\ref{eq:Vcw}) and~(\ref{eq:Vct}), as well as thermal corrections~(\ref{eq:VT}):
\begin{equation}
    \label{eq:Veff}
    V_{\text{eff}} (h, H, s, T) = V_{\text{cl}} (h, H, s) + V_{\text{CW}} (h, H, s, T) + \delta V(h, H, s) + V_{\text{T}} (h, H, s, T).
\end{equation}

%%%%%%%%%%%%%%%%%%%%%%%%%%%%%%%%%%%%%%%%%%%%%%%%%%%%%%%%%%

\section{Constraints}
\label{sec:constraints}

%%%%%%%%%%%%%%%%%%%%%%%%%%%%%%%%%%%%%%%%%%%%%%%%%%%%%%%%%%
\subsection{Perturbativity}

We require the Feynman rule vertex factors of quartic interactions to be less than $4 \pi$ in absolute value to ensure that the one-loop contributions be smaller than tree-level ones \cite{Lerner:2009xg}. This yields the conditions
\begin{equation}
\begin{aligned}
  \abs{\lambda_{1}} &< \frac{2 \pi}{3}, 
  &
  \abs{\lambda_{2}} &< \pi, 
  &
  \abs{\lambda_{3}} &< 4 \pi, 
  &
  \abs{\lambda_{4}} &< 4 \sqrt{2} \pi, 
  &
  \abs{\lambda_{3} + \lambda_{4}} &< 4 \pi, 
  \\
  \abs{\lambda_{S}} &< \pi, 
  &
  \abs{\lambda_{S1}} &< 4 \pi, 
  &
  \abs{\lambda_{S2}} &< 4 \pi, 
  &
  \abs{\lambda_{S12}} &< 4 \pi.
\end{aligned}
\end{equation}

%%%%%%%%%%%%%%%%%%%%%%%%%%%%%%%%%%%%%%%%%%%%%%%%%%%%%%%%%%
\subsection{Unitarity}

The cross section of $2 \to 2$ scattering processes $s_{1} s_{2} \to s_{3} s_{4}$ of scalars $s_{i}$ can be expanded in the partial wave decomposition as
\begin{equation}
  \sigma = \frac{16 \pi}{s} \sum_{l = 1}^{\infty} (2 l + 1) \abs{a_{l}(s)}^{2},
\end{equation}
where $s$ is the Mandelstam variable and $a_{l}$ are the partial wave coefficients with the angular momenta $l$. The optical theorem imposes on $a_{l}$ the unitarity bound
\begin{equation}
  \abs{\Re (a_{l})} < \frac{1}{2}.
  \label{eq:unit:bound}
\end{equation}
In the high energy limit, the $s$-wave amplitude $a_{0}(s)$ is dominated by contact terms because the $s$-, $t$- and $u$-channel processes are suppressed by the scattering energy. Therefore, in the high energy limit, $a_{0}(s)$ is completely determined by the quartic couplings of the scalar potential. The constraint \eqref{eq:unit:bound} places an upper bound on the eigenvalues of the $2 \to 2$ scalar scattering matrix. The unitarity bounds of the 2HDM were first studied in \cite{Kanemura:1993hm,Akeroyd:2000wc}. For the $\mathbb{Z}_{3}$ scalar quartic couplings, the unitarity bounds are given by\footnote{We correct the result in~\cite{Belanger:2014bga}, where an overall factor of $2$ is missing from  the scattering matrix and the coupling $\lambda_S$ should be replaced with $2\lambda_S$.}
\begin{align}
\vert\lambda_1\vert&\leq 4\pi, \\
\vert\lambda_2\vert&\leq 4\pi, \\
\vert\lambda_3\vert&\leq 8\pi, \\
\vert\lambda_3-\lambda_4\vert&\leq 8\pi, \\
\vert\lambda_3+\lambda_4\vert&\leq 8\pi, \\
\biggr\vert\lambda_1+\lambda_2\pm\sqrt{(\lambda_1-\lambda_2)^2+\lambda_4^2}\biggr\vert&\leq 8\pi, \\
\biggr\vert\lambda_3+2\lambda_4+2\lambda_S\pm\sqrt{(\lambda_3+2\lambda_4-2\lambda_S)^2+4\lambda_{S12}^2}\biggr\vert&\leq 16\pi, \\
\vert\lambda_{S1}\vert&\leq 8\pi, \\
\vert\lambda_{S2}\vert&\leq 8\pi, \\
\biggr\vert\lambda_{S1}+\lambda_{S2}\pm\sqrt{(\lambda_{S1}-\lambda_{S2})^2+4\lambda_{S12}^2}\biggr\vert&\leq 16\pi
\end{align}
and $\abs{\Lambda_{i}} \leq 1/2$, where the three remaining eigenvalues $\Lambda_{i}$ with $i = 1,2,3$ of the scattering matrix are given by the solution of the cubic equation
\begin{equation}
\begin{split}
  &x^3 + \frac{1}{8\pi} x^2 (3 \lambda_1 + 3 \lambda_2 + 2 \lambda_S) + \frac{1}{256\pi^2} x 
  (36 \lambda_1 \lambda_2 - 4 \lambda_3^2 - 4 \lambda_3 \lambda_4 - \lambda_4^2 + 24 \lambda_1 \lambda_S 
  + 24 \lambda_2 \lambda_S 
  \\
  &- 2 \lambda_{S1}^2 - 2 \lambda_{S2}^2) + \frac{1}{1024\pi^3} (36 \lambda_1 \lambda_2 \lambda_S 
  - 4 \lambda_3^2 \lambda_S - 4 \lambda_3 \lambda_4 \lambda_S - \lambda_4^2 \lambda_S 
  - 3 \lambda_2 \lambda_{S1}^2 + 2 \lambda_3 \lambda_{S1} \lambda_{S2} 
  \\
  &+ \lambda_4 \lambda_{S1} \lambda_{S2} - 3 \lambda_1 \lambda_{S2}^2) = 0.
\end{split}
\end{equation}

%%%%%%%%%%%%%%%%%%%%%%%%%%%%%%%%%%%%%%%%%%%%%%%%%%%%%%%%%%
\subsection{Bounded-from-below constraints and globality of the electroweak vacuum}

In order to be physical, the potential has to be bounded from below in the limit of large field values. In this limit, the dimensionful quadratic and cubic terms can be neglected --- it suffices to consider only the quartic part of the potential~\eqref{eq:V:c}. The analytical necessary and sufficient bounded-from-below constraints were derived in Ref.~\cite{Kannike:2016fmd}. We only sketch the derivation and refer the reader to that paper for details and ancillary \textit{Mathematica} code.

The potential~\eqref{eq:V:c}, using the parameterisation
\begin{equation}
  \abs{H_{1}}^{2} = h_{1}^{2}, \quad \abs{H_{2}}^{2} = h_{2}^{2},
  \quad \hc{H_{1}} H_{2} = h_{1} h_{2} \rho e^{i \phi}, \quad S = s e^{i \phi_{S}},
  \label{eq:V:2HDM:param}
\end{equation}
where $0 \leq \rho \leq 1$ as implied by the Cauchy inequality $0 \leqslant \abs{\hc{H_{1}} H_{2}} \leqslant \abs{H_{1}} \abs{H_{2}}$, and $0 \leq \phi, \phi_{S} < 2 \pi$, takes the form
\begin{equation}
\begin{split}
  V &= \lambda_{1} h_{1}^{4} + \lambda_{2} h_{2}^{4} + \lambda_{3} h_{1}^{2} h_{2}^{2}
  + \lambda_{4} \rho^{2} h_{1}^{2} h_{2}^{2} + \lambda_{S} s^{4}
  \\
  &+ \lambda_{S1} s^{2} h_{1}^{2}
  + \lambda_{S2} s^{2} h_{2}^{2} - \abs{\lambda_{S12}} \rho s^{2} h_{1} h_{2},
\end{split}
\label{eq:V:Z3:SIID}
\end{equation}
where we have minimised $\cos (\phi + 2 \phi_{S} + \phi_{\lambda_{S12}}) = -1$ so $\lambda_{S12} = -\abs{\lambda_{S12}}$ without loss of generality. We have to minimise the potential with respect to $\rho$, also taking into account separately the ends $\rho = 0$ and $\rho = 1$ of the interval.
The conditions for $s = 0$ or $\rho = 0$ are given by
\begin{align}
  \lambda_{1} &> 0, 
  &
  \lambda_{2} > 0,
  \qquad \qquad \qquad \qquad \qquad 
  \lambda_{S} &> 0,
  \notag
  \\
  \bar{\lambda}_{3} \equiv \lambda_{3} + 2 \sqrt{\lambda_{1} \lambda_{2}} &> 0, 
  \span
  \bar{\lambda}_{34} \equiv  \lambda_{3} + \lambda_{4} + 2 \sqrt{\lambda_{1} \lambda_{2}} &> 0,
  \notag
  \\
  \bar{\lambda}_{S1} \equiv \lambda_{S1} + 2 \sqrt{\lambda_{S} \lambda_{1}} &> 0, 
  &
  \bar{\lambda}_{S2} \equiv \lambda_{S2} + 2 \sqrt{\lambda_{S} \lambda_{2}} &> 0 
  \notag
  \\
  \span &
  \sqrt{\lambda_{S}} \lambda_{3} + \sqrt{\lambda_{1}} \lambda_{S2} + \sqrt{\lambda_{2}} \lambda_{S1} 
  + \sqrt{\lambda_{S} \lambda_{1} \lambda_{2}} 
  + \sqrt{\bar{\lambda}_{S1} \bar{\lambda}_{S2} \bar{\lambda}_{3}} &> 0.
  \label{eq:BfB:rho:0:s:0}
\end{align}
Another necessary but not sufficient (since it is calculated with $\rho =1$, but ignoring the $\lambda_{S12}$ term) condition can be given by 
\begin{equation}
   \sqrt{\lambda_{S}} (\lambda_{3} + \lambda_{4}) + \sqrt{\lambda_{1}} \lambda_{S2} + \sqrt{\lambda_{2}} \lambda_{S1} 
  + \sqrt{\lambda_{S} \lambda_{1} \lambda_{2}} 
  + \sqrt{\bar{\lambda}_{S1} \bar{\lambda}_{S2} \bar{\lambda}_{34}} > 0.
  \label{eq:BfB:rho:1:laS12:0}
\end{equation}

We minimise the potential with respect to $h_{1}$, $h_{2}$, $s$ and $\rho$ with the fields lying on a sphere, enforced by a Lagrange multiplier $\lambda$. The minimisation equations with respect to the fields and $\lambda$ are
\begin{equation}
\begin{split}
  4 \lambda_{1} h_{1}^{3}  + 2 (\lambda_{3} + \lambda_{4} \rho^{2}) h_{1} h_{2}^{2} 
  + 2 \lambda_{S1} h_{1} s^{2} - \abs{\lambda_{S12}} \rho h_{2} s^{2}
  &= \lambda h_{1},
  \\
  4 \lambda_{2} h_{2}^{3}  + 2 (\lambda_{3} + \lambda_{4} \rho^{2}) h_{1}^{2} h_{2} 
  + 2 \lambda_{S2} h_{2} s^{2} - \abs{\lambda_{S12}} \rho h_{1} s^{2}
  &= \lambda h_{2},
  \\
  s \left( 4 \lambda_{S} s^{2} + 2 \lambda_{S1} h_{1}^{2} + 2 \lambda_{S2} h_{2}^{2} 
  - 2 \abs{\lambda_{S12}} \rho h_{1} h_{2} \right) &= \lambda s,
  \\
  h_{1}^{2} + h_{2}^{2} + s^{2} &= 1.
\end{split}
\label{eq:Bfb:eq:sphere}
\end{equation}
and the minimisation equation with respect to $\rho$ is
\begin{equation}
    h_{1} h_{2} \left( 2 \rho \lambda_{4} h_{1} h_{2} - \abs{\lambda_{S12}} s^{2} \right) = 0,
\label{eq:Bfb:eq:rho}
\end{equation}
The equations \eqref{eq:Bfb:eq:sphere} and \eqref{eq:Bfb:eq:rho}, solved together, have an analytical solution. The bounded-from-below conditions for $0 < \rho < 1$ are then given by
\begin{equation}
  0 < h_{1}^{2} < 1 \land 0 < h_{2}^{2} < 1 
  \land 0 < s^{2} < 1 
  \land 0 < \rho^{2} < 1 \implies V_{\text{min}} > 0,
\label{eq:bfb:implication}
\end{equation}
where $V_{\text{min}}$, the value of minimum at the solution, is proportional to the Lagrange multiplier $\lambda$. Note that $p \implies q$ is equivalent to $\lnot p \lor q$.
In addition, the equations \eqref{eq:Bfb:eq:sphere} need also to be solved with $\rho = 1$, in which case the conditions are given by\footnote{Another form of these conditions is given in Ref.~\cite{Kannike:2016fmd}.}
 \begin{equation}
  0 < h_{1} < 1 \land 0 < h_{2} < 1 
  \land 0 < s < 1 
 \implies V_{\text{min}} > 0.
\label{eq:bfb:implication:rho:1}
\end{equation}

The necessary and sufficient bounded-from-below conditions are then given by Eqs.~\eqref{eq:BfB:rho:0:s:0}, \eqref{eq:bfb:implication} and \eqref{eq:bfb:implication:rho:1}. Eq.~\eqref{eq:BfB:rho:1:laS12:0} can be used for faster elimination of invalid points.

Once the bounded-from-below conditions are satisfied, the potential is guaranteed to have a finite global minimum solution. Although it is possible that our EW vacuum, not coinciding with the global one, is metastable, the gain in the parameter space is expected to be marginal (at least in the singlet-like case \cite{Hektor:2019ote}). For that reason, we simply demand that the EW vacuum be the global one.

%%%%%%%%%%%%%%%%%%%%%%%%%%%%%%%%%%%%%%%%%%%%%%%%%%%%%%%%%%
\subsection{Collider constraints}

The LEP collider measured to high precision the decay widths of the $Z$ and $W$ bosons, which are compatible with the SM. Neglecting the small mixing between the singlet and doublet, we require that the particle masses satisfy~\cite{Cao:2007rm}
\begin{equation}
    M_{x_{2}} + M_{H^\pm} > M_{W}, \quad 2 M_{x_{2}} > M_{Z}, \quad 2 M_{H^\pm} > M_{Z}.
\end{equation}
LEP searches for new neutral final states further exclude a range of masses~\cite{Lundstrom:2008ai}, thereby forcing
\begin{equation}
    m_H > 80~\mathrm{GeV}, \quad m_A > 100~\mathrm{GeV} \quad\text{or} \quad m_A - m_H < 8~\mathrm{GeV},
\label{eq:LEP}
\end{equation}
in addition to
\begin{equation}
    m_{H^\pm} > 70~\mathrm{GeV}
\end{equation}
due to searches for charged scalar pair production~\cite{Pierce:2007ut}. But since there is no splitting between $m_{H}$ and $m_{A}$, the constraint \eqref{eq:LEP} is always satisfied.

Similarly, if $M_{x_{1}} < M_{h}/2$ or $M_{x_{2}} < M_{h}/2$, the Higgs boson can decay into dark sector.
This is constrained by measurements of the Higgs boson invisible branching ratio $\text{BR}_{\text{inv}} = \Gamma_{h \to x_{1} x_{1}} / (\Gamma_{h \to \text{SM}} + \Gamma_{h \to x_{1} x_{1}})$. Latest measurements by the CMS experiment at the LHC find that $\text{BR}_{h \to \text{inv}}< 0.15$ \cite{CMS:2023sdw}, while the ATLAS experiment measures $\text{BR}_{h \to \text{inv}}< 0.11$ \cite{ATLAS:2023tkt}, of which we will require the stricter ATLAS constraint.

We also use the micrOMEGAs package to compute the $h \to \gamma\gamma$ decay rate (usually similar to the IDM one \cite{Swiezewska:2012eh,Krawczyk:2013jta}) and require it to be within the value measured at the CMS \cite{CMS:2021kom}: $\mu_{h \to \gamma\gamma} = 1.12 \pm 0.09$.

%%%%%%%%%%%%%%%%%%%%%%%%%%%%%%%%%%%%%%%%%%%%%%%%%%%%%%%%%%
\subsection{Electroweak precision parameters}
\label{sec:EWPT}

The measurements of electroweak precision data at the LEP put strong constraints on physics beyond the SM. The latest electroweak fit by the Gfitter group \cite{Haller:2018nnx} gives for the oblique parameters $S$ and $T$ (with $U = 0$) the central values
\begin{equation}
  S = 0.04 \pm 0.08, \quad T = 0.08 \pm 0.14,
\end{equation}
with a correlation coefficient of $+0.92$.

To calculate electroweak precision parameters $S$ and $T$, we use the results for general models with doublets and singlets \cite{Grimus:2008nb,Grimus:2007if}. The usual loop functions are defined as
\begin{equation}
  F(I, J) = \frac{I + J}{2} - \frac{I J}{I - J} \ln{\frac{I}{J}},
\end{equation}
with $F(I,I) = 0$ in the limit of $J \to I$, and
\begin{equation}
\begin{split}
  G \left( I, J, Q \right) &=
  - \frac{16}{3} + \frac{5 \left( I + J \right)}{Q}
  - \frac{2 \left( I - J \right)^2}{Q^2}
  \\
  &+ \frac{3}{Q} \left[ \frac{I^2 + J^2}{I - J}
- \frac{I^2 - J^2}{Q}
+ \frac{\left( I - J \right)^3}{3 Q^2} \right]
\ln{\frac{I}{J}}
+ \frac{r}{Q^3}\, f \left( t, r \right),
\end{split}
\end{equation}
where
\begin{equation}
  t \equiv I + J - Q 
  \quad \text{and} \quad
  r \equiv Q^2 - 2 Q (I + J) + (I - J)^2,
\end{equation}
\begin{equation}
\begin{split}
  f \left( t, r \right) \equiv \left\{ \begin{array}{lcl}
{
\sqrt{r}\, \ln{\left| \frac{t - \sqrt{r}}{t + \sqrt{r}} \right|}
} & \Leftarrow r > 0,
\\*[3mm]
0 & \Leftarrow r = 0,
\\*[2mm]
{
2\, \sqrt{-r}\, \arctan{\frac{\sqrt{-r}}{t}}
} & \Leftarrow  r < 0.
\end{array} \right.
\end{split}
\end{equation}
The beyond-the-SM contributions to the $S$ and $T$ parameters are given by
\begin{equation}
\begin{split}
  \Delta S &= \frac{1}{24 \pi} \left[ (2 s_{W}^2 -1)^2 \, G(M_{H^{\pm}}^{2},M_{H^{\pm}}^{2},M_{Z}^{2}) 
  + \cos^{4} \theta \, G(M_{x_{2}}^{2},M_{x_{2}}^{2},M_{Z}^{2}) \right. \\ 
  & + 2 \sin^{2} \theta \cos^{2} \theta \, G(M_{x_{1}}^{2},M_{x_{2}}^{2},M_{Z}^{2}) 
  + \sin^{4} \theta \, G(M_{x_{1}}^{2},M_{x_{1}}^{2},M_{Z}^{2}) \\
  & \left. + 2 \sin^{2} \theta \ln M_{x_{1}}^{2} + 2 \cos^{2} \theta \, \ln M_{x_{2}}^{2} 
  - 2 \ln M_{H^{\pm}}^{2} \right] \\
  &\approx \frac{1}{24 \pi}
  \left[ (2 s_{W}^2 -1)^2 \, G(M_{H^{\pm}}^{2},M_{H^{\pm}}^{2},M_{Z}^{2}) 
  + G(M_{x_{2}}^{2},M_{x_{2}}^{2},M_{Z}^{2}) \right. \\
  &\left. + 2 \ln M_{x_{2}}^{2} - 2 \ln M_{H^{\pm}}^{2} \right],
\end{split}
\end{equation}
and
\begin{equation}
\begin{split}
  \Delta T &= \frac{1}{16 \pi^{2} \alpha v^{2}} \left[ \sin^{2} \theta \, F(M_{H^{\pm}}^{2}, M_{x_{1}}^{2})  
  + \cos^{2} \theta \, F(M_{H^{\pm}}^{2}, M_{x_{2}}^{2})  \right. \\
  &\left. - \sin^{2} \theta \cos^{2} \theta \, F(M_{x_{1}}^{2}, M_{x_{2}}^{2})  \right] \\
  &\approx \frac{1}{16 \pi^{2} \alpha v^{2}} F(M_{H^{\pm}}^{2}, M_{x_{2}}^{2}),
\end{split}
\end{equation}
where the approximations are given for a small mixing angle.
The electroweak precision constraints mean that $M_{H^{\pm}}$ cannot be too different from $M_{x_{2}}$. We require the $S$ and $T$ parameters to fall within the 95\% C.L. ellipse.

%%%%%%%%%%%%%%%%%%%%%%%%%%%%%%%%%%%%%%%%%%%%%%%%%%%%%%%%%%
\section{Relic density}
\label{sec:relic}

%Since the $\lambda_{5}$ term in the potential is forbidden by the $\mathbb{Z}_{3}$ symmetry, the 
The cosmological abundance of DM measured by the Planck satellite $\Omega_c h^2 = 0.120 \pm 0.001$ \cite{Planck:2018vyg} significantly constrains the allowed parameter space of the model.
The model that we study contains a non-trivial dark sector that is comprised of several particles with $\mathbb{Z}_{3}$ quantum numbers: $x_{1}$, $x_{2}$ and $H^{\pm}$.
Among them only the singlet-like $x_{1}$ can play the role of DM, since the physical states that originate from the inert doublet $H_{2}$ are too strongly coupled to the SM fields to satisfy the constraints and requirements for the DM candidate. Nevertheless, the processes that involve other dark sector particles in the early Universe can leave an imprint on the relic abundance and generally has to be included in the calculation of DM density evolution. 
In similar models with semi-annihilation \cite{Belanger:2012vp,Belanger:2014bga,Belanger:2021lwd,Belanger:2020hyh}, the relic density is obtained under the assumption that all the particles were initially in thermal equilibrium with the SM plasma and furthermore that the kinetic equilibrium is maintained throughout the whole DM production stage. In this case the density of DM can be obtained by solving the set of Boltzmann equations for the densities of different particles in the dark sector, e.g. using available numerical solvers like micrOMEGAs \cite{Belanger:2018ccd,Belanger:2020gnr,Alguero:2022inz} that can work with multi-component DM models \cite{Belanger:2013oya}. If the processes of DM scattering on the SM plasma particles are not very efficient, the kinetic decoupling can occur before the freeze-out. In this case semi-annihilation processes can redistribute the energy of DM particles and modify its temperature,\footnote{Here we consider a simplified picture of DM evolution, which holds on the assumption that the DM distribution has an equilibrium \textit{shape}, leading to coupled Boltzmann equations (cBE) for the moments of the distribution function. In general one has to solve the full Boltzmann equation (fBE) for the DM distribution function as described in Ref.~\cite{Binder:2021bmg}, however in this work we follow the assumption above.} which has an impact on the rates of annihilation processes.   
Although this assumption is often justified or introduces very small corrections to the relic density, in some regimes the process that 
defines the dynamics of the freeze out can be strongly velocity-dependent, e.g. due to a resonance or $s$-wave suppression, so that its rate becomes sensitive to the temperature of DM. The density evolution of $\mathbb{Z}_{3}$ scalar singlet DM with semi-annihilation beyond the assumption of kinetic equilibrium was studied in Ref.~\cite{Hektor:2019ote}. In this work we use the same method to calculate the relic density, but extend it to the multi-component case of the model that we study and include additional interactions present in this model.       

In order to properly take into account the departure from kinetic equilibrium of the DM particles and its effect on the relic abundances, we follow the formalism of \cite{Binder:2017rgn} and the numerical implementation as presented in the DRAKE code release paper \cite{Binder:2021bmg}, with the addition of semi-annihilation collision term as discussed in \cite{Hektor:2019ote}. %Additionally, the model at hand involves a dark sector with three particles whose dynamics can be relevant for the evolution of the DM relic abundance and therefore below we briefly review the adopted formalism extended to multi-component scenarios. 
We focus on the cBE method, i.e. coupled Boltzmann equations for the $0^{\text{th}}$ and $2^{\text{nd}}$ moment of the distribution function, describing the evolution of number density and temperature, respectively. In doing so we neglect potential corrections from the departure of the thermal shape of the DM distribution.\footnote{These are expected to be sub-leading and their proper calculation would require much more complicated calculation involving not only self-interaction processes \cite{Hryczuk:2022gay}, but also an implementation of the semi-annihilation collision term at the phase-space level, which is considerably more CPU expensive.} Below we briefly review the adopted formalism.  

The formulation of the system of Boltzmann equations that we solve originates from the master equation governing the evolution of the phase-space density $f_i=f_i(t,\mathbf{p})$ of particle $i$ in an expanding Friedmann-Lemaître-Robertson-Walker universe: 
\begin{equation}
  \label{eq:BE}
  E\left(\partial_t-H\mathbf{p}\cdot\nabla_\mathbf{p}\right)f_i=C[f_i]\,,
\end{equation}
where $H$ is the Hubble parameter and the collision term $C[f_i]$ contains all possible binary and non-binary interactions between dark sector and/or SM particles. Taking the first two moments of the above equation, i.e. integrating it over $g_i\int d^3p/(2\pi)^3/E$ and $g_i\int d^3p/(2\pi)^3$ $\mathbf{p}^2/E^2$,
respectively, with $g_i$ being the number of spin degrees of freedom, and introducing new variables $Y\equiv n/s$ and $ y\equiv \frac{m_\chi}{3 s^{2/3}}
\left\langle \frac{\mathbf{p}^2}{E} \right\rangle$ with $x=m_{x_1}/T$, one finds:
\begin{align}
\frac{Y_i'}{Y_i} &= \frac{m_i}{x \tilde H}C_0\,, \label{Yfinal}\\
\frac{y_i'}{y_i} &= \frac{m_i }{x \tilde H} C_2 - \frac{Y_i'}{Y_i} 
+\frac{H}{x\tilde H}
\frac{g_i}{3n_i T_i}\int \frac{d^3p}{(2\pi)^3}\,\frac{\mathbf{p}^4}{E^3} f_i(\mathbf{p})\,,\label{yfinal}
%\frac{m_\chi}{yYs^{5/3}} \frac{g_\chi}{6\pi^2} \!\!\int\!\! dp\, \frac{p^{6}}{E^3}f_\chi, \label{yfinal}
\end{align}
where we introduce the moments of the collision term as
\begin{align}
m_i n_i C_0&\equiv  g_i \int \frac{d^3p}{(2\pi)^3E}\, C[f_i]\,, \label{eq:C0}\\
m_i n_i \left\langle \frac{\mathbf{p}^2}{E} \right\rangle C_2 &\equiv  g_i \int \frac{d^3p}{(2\pi)^3E} \frac{\mathbf{p}^2}{E}\, C[f_i]\, \label{eq:C2}.
\end{align}
This set of cBE \eqref{Yfinal} and \eqref{yfinal} is not closed in terms of higher moments unless some assumption is made on the form of the distribution function. A well motivated one is to make an ansatz
\begin{eqnarray}
f_i(E_i,T_i)=\frac{n_i(T_i)}{n_i^{\text{eq}}(T_i)} e^{-E_i/T_i },
\label{eq:ansatz}
\end{eqnarray}
which enforces the Maxwell-Boltzmann shape, but with an arbitrary temperature $T_i$. In the scenario studied in this work we have $i\in \{x_1,x_2,H^\pm\}$, but we will make a simplifying assumption that only the DM particle, $x_1$, has a population which departs from kinetic equilibrium, while all three dark sector states can depart from chemical equilibrium. This assumption is justified as $x_2$ and $H^\pm$ are relatively tightly coupled to the thermal bath, leading to their kinetic decoupling happening well after all the freeze-out processes take place, while in the regions of dominant semi-annihilation the $x_1$ has suppressed elastic scatterings and thus can undergo an early kinetic decoupling.

The general expression for the collision term for $i$-th particle due to binary collisions reads
\begin{eqnarray}
\label{eq:Cann}
C = \frac{1}{2g_i}\sum_{j\{kl\}}  \frac{1}{(1+\delta_{kl})} &&\!\!\!\!\!\!\!\!~\int\! d\Pi_j d\Pi_k d\Pi_l (2\pi)^4\delta^{(4)}(p_i+p_j-p_k-p_l)   \MM_{ij \leftrightarrow kl} \nonumber \\ 
&&\times \left[f_k f_l(1\pm f_i)(1\pm f_j)-f_i f_j(1\pm f_k)(1\pm f_l)\right] \, ,
\end{eqnarray}
where the sum goes over all possible $j$ and two-particle $\{kl\}$ states, with and without the DM, and $\MM$ is the amplitude squared summed over initial and final internal d.o.f. It is understood that $\{kl\}=\{lk\}$ and it appears in the sum only once. The symmetry factor $(1+\delta_{kl})^{-1}$  is related to the exchange of identical particles in  the final state. This formula incorporates annihilations,  semi-annihilations and inelastic scatterings. Following \cite{Bringmann:2006mu} and \cite{Binder:2016pnr}, we treat the elastic scatterings in the Fokker-Planck approximation, effectively involving expansion in small momentum transfer. 

Including also all possible decays of $x_2$ and $H^\pm$, and neglecting the quantum degeneracy effects in all the cases that we consider, such that $(1 \pm f) \approx 1$,  the cBE system for the $i$-th particle has the form:
\begin{align}
\frac{Y_i'}{Y_i} =& -\frac{s}{x \tilde H}\frac{1}{Y_i} \Biggl\{ \sum_{j\{kl\}} \left(Y_i Y_j I_{ij\rightarrow kl} - Y_k Y_l \frac{(1+\delta_{ij})}{(1+\delta_{kl})} I_{kl\rightarrow ij}  \right) \nonumber\\
 &- \frac{m_j^2 T}{3\pi^2 s^2}K_1\left(\frac{m_j}{T}\right)\left(\frac{Y_j}{Y_j^{\text{eq}}}-\frac{Y_i}{Y_i^{\text{eq}}}\right)\sum_{\{kl\}}  \frac{\Gamma_{j\to i k l}}{(1+\delta_{kl})} + (j \leftrightarrow i) \Biggr\} \, , \label{eq:YcBE} \\
\frac{y'}{y} +  \frac{Y_{x_1}'}{Y_{x_1}} =& -\frac{s}{x \tilde H} \frac{1}{Y_{x_1}}\sum_{j\{kl\}} \left(  Y_{x_1} Y_j J_{{x_1} j\rightarrow kl}  - Y_k Y_l \frac{(1+\delta_{{x_1} j})}{(1+\delta_{kl})} \frac{\sqrt{T_k T_l}}{T_{x_1}} K_{kl\rightarrow {x_1} j}\right) \nonumber \\
& + \frac{s}{x \tilde H} \frac{1}{Y_{x_1}}\sum_{j\{kl\}} Y_j \frac{D_{j\to i k l}}{(1+\delta_{kl})}  + \frac{\gamma(x)}{x \tilde H} \left( \frac{y^{\text{eq}}}{y}-1\right)  \nonumber \\
&+\frac{H}{x\tilde H}
\frac{g_{x_1}}{3n_{x_1} T_{x_1}}\frac{Y_{x_1}}{Y_{x_1}^{\text{eq}}}\int \frac{d^3p}{(2\pi)^3}\,\frac{\mathbf{p}^4}{E^3} f_{x_1}^{\text{eq}}(\mathbf{p}) \, ,
\end{align}
where $\tilde H\equiv H/\left[1+ \tilde g(x)\right]$ and $\tilde g \equiv \frac{1}{3} \frac{T}{g^s_{\rm\rm{eff}}}\frac{ d g^s_{\rm\rm{eff}}}{d T}$ with $g^s_{\rm\rm{eff}}$ being the number of entropy degrees of freedom, the sum goes over all possible $j$ states and $\{kl\}$ two-particle states, $K_1$ is the modified Bessel function of order~1, $\Gamma_{j \rightarrow ikl}$ stands for the decay width of the $j$-particle into $ikl$ states and the rates are defined below.\footnote{Note that there are no additional factors related to the number of particles being created or destroyed in a given process, as those come up automatically when one starts from the expression \eqref{eq:Cann}. In particular, for semi-annihilation one needs to 
sum over the processes in which the particle under consideration appears in the initial and in the final states, e.g., $ij\to \{kl\}=x_1 x_1\to \{X_1 Z\}$ as well as $ij\to \{kl\}=x_1 Z\to \{X_1 X_1\}$. Accounting for the symmetry factor in the general collision term this leads to the well-known form of the Boltzmann equation, where the semi-annihilation term has a factor $1/2$ compared to the annihilation term (see the notes inside Ref.~\cite{DK}).} This is a set of 4 coupled ordinary differential equations that describe the evolution of comoving number densities of $x_1$, $x_2$ and $H^\pm$, as well as the temperature of $x_1$.

The evolution of the above system is governed by the interaction rates in the form of the momentum transfer rate $\gamma(x)$, as defined in \cite{Binder:2017rgn}, and integrals $I,J,K$ defined as follows. The number-changing processes are described by the following integral\footnote{This is a generalization of the thermally averaged cross section for the case of \mbox{(co-)(semi-)annihilating} particles with different temperatures. Indeed, for $T_i = T_j \equiv T'$ this expression is the same as in
Ref.~\cite{Edsjo:1997bg}. Following their evaluation one finds
\begin{equation}
I_{ij\rightarrow kl}(T') = \frac{1}{n^{\text{eq}}_i n^{\text{eq}}_j} \frac{T'}{32\pi^4}\int_{(m_i+m_j)^2}^\infty \! ds\, g_i g_j p_{ij} W_{ij\rightarrow kl} K_1\left(\frac{\sqrt{s}}{T'}\right) = \langle \sigma v_{\rm Mol} \rangle_{ij\to kl},
\end{equation}
where we used $p_{ij}=  \sqrt{s-(m_i+m_j)^2}\sqrt{s-(m_i-m_j)^2}/(2\sqrt{s})$.}
\begin{equation}
I_{ij\rightarrow kl} =\frac{1}{n^{\text{eq}}_i n^{\text{eq}}_j} \int d\Pi_i d\Pi_j g_i g_j W_{ij\rightarrow kl} f^{\text{eq}}_i (T_i) f^{\text{eq}}_j (T_j),
\end{equation}
through the use of the dimensionless invariant rate \cite{Edsjo:1997bg},
\begin{equation}
W_{ij\rightarrow kl}(s) \equiv \frac{1}{g_i g_j (1+\delta_{kl})}\int d\Pi_k d\Pi_l (2\pi)^4\diracd{p_i}{p_j}{p_k}{p_l}  \MM_{ij \leftrightarrow kl}
\end{equation}
that is related to the cross section via $W_{ij\rightarrow kl} = 4 p_{ij} \sqrt{s} \sigma_{ij}.$ %= 4 E_i E_j \sigma_{ij} v_{ij}$. 
The integrals involved in the temperature evolution,
\begin{align}
J_{ij\rightarrow kl} &= \frac{1}{T_i n^{\text{eq}}_i n^{\text{eq}}_j}\int d\Pi_i d\Pi_j  \frac{p_i^2}{3E_i}  g_i g_j  W_{ij \to kl}  f^{\text{eq}}_i (T_i) f^{\text{eq}}_j (T_j),
\\
K_{kl\rightarrow ij} &= \frac{1}{\sqrt{T_k T_l}n^{\text{eq}}_k n^{\text{eq}}_l}\frac{1}{(1+\delta_{ij})}\int d\Pi_i d\Pi_j d\Pi_k d\Pi_l (2\pi)^4\diracd{p_i}{p_j}{p_k}{p_l}  \frac{p_i^2}{3E_i}
\notag
\\
&\times \MM_{ij \leftrightarrow kl}  f^{\text{eq}}_k(T_k) f^{\text{eq}}_l (T_l), 
\\
D_{j \rightarrow ikl} &= \frac{1}{3T_i n^{\text{eq}}_j}\int d\Pi_i d\Pi_j d\Pi_k d\Pi_l (2\pi)^4\delta(p_i-p_j-p_k-p_l)  \frac{p_i^2}{E_i}\MM_{j \leftrightarrow ikl}  f^{\text{eq}}_j (T_j) ,
\end{align}
describe the change in energy due to forward processes (annihilations), backward processes (inverse annihilations) and decays, respectively.

Numerical solution of the above cBE system is relatively CPU expensive, in large part due to necessity of calculating a large number of rates.\footnote{Note that in contrast to annihilation processes, the inelastic scatterings and semi-annihilations cannot be always included with the use of inclusive cross sections, but rather one has to perform the thermal average for every exclusive process separately.} To render the task manageable, the numerical procedure we follow is to first determine for each of the parameter point if the full cBE system is needed or if the standard approach assuming kinetic equilibrium for the DM particle should suffice. For that we impose a rather conservative condition that if $\gamma(x)/H(x)|_{x=100} \geq 10$, then for the whole period of chemical freeze-out the elastic scatterings are assumed to be efficient enough to keep the DM in kinetic equilibrium, even in the presence of other disruptive processes like decay of a heavier dark sector state, and the assumption of $T_{x_1}=T$ is well justified allowing for calculations using only the Eq.~\eqref{eq:YcBE}. Conversely, if the scattering rate is smaller, $\gamma(x)/H(x)|_{x=100} < 10$, then the full cBE system is solved.

The needed rates were calculated using FeynCalc \cite{Shtabovenko:2020gxv} for the matrix elements and using the DRAKE thermal average routines. On top of that the decay widths were utilized from the micrOMEGAs calculation. The latter was also used to include in the number density equation typically sub-dominant, but in some cases important annihilations to final states including virtual $Z$ or $W$ bosons, where we followed the procedure as described in the micrOMEGAs manual \cite{Belanger:2013oya}.
Heavier dark sector states were assumed to take part in co-annihilations and inelastic scattering processes if their mass did not exceed 50\% of the DM state and the binary collision rates were truncated for processes that contribute less than 1\% at the approximate time of freeze-out.
 
\begin{figure}[h!]
\begin{center}
  \includegraphics[width=0.51\linewidth]{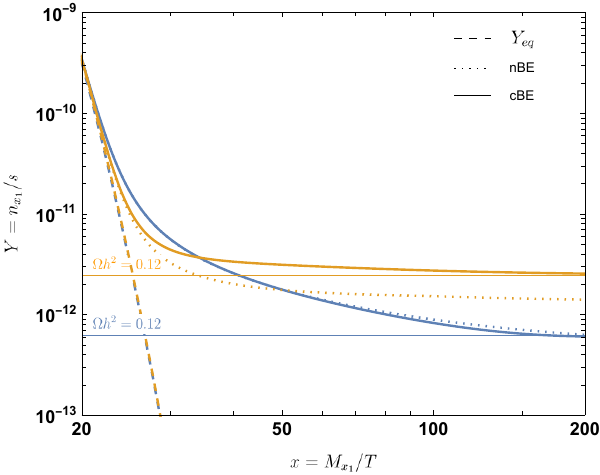}
    \includegraphics[width=0.48\linewidth]{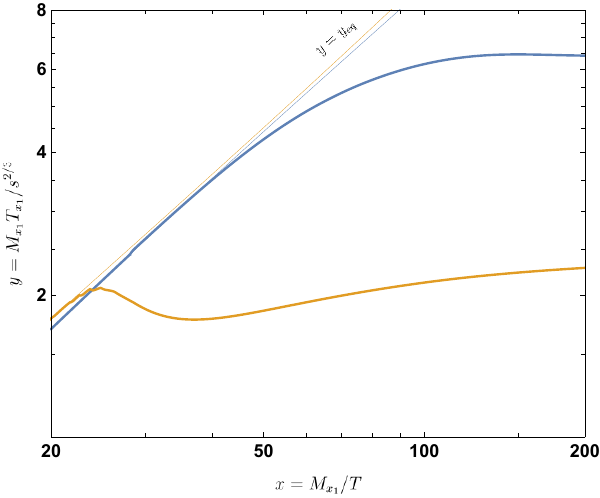}
\caption{An example evolution of the yield $Y$ \textit{(left)} and the temperature parameter $y$ \textit{(right)} for two benchmark points with parameters: 
$M_{x_1}=691.38$ GeV, $M_{x_2}= 1388.03$ GeV, $M_{H^+}=1388.03$ GeV, $\sin\theta=0.001264$, $\mu^{''}_S=3943.34$ GeV, $\lambda_{S12}=\lambda_{S1}=\lambda_{S2}=0$, $\lambda_{S}=\lambda_2=\pi$, $\lambda_3=0$ (blue) 
and
$M_{x_1}=175.31$ GeV, $M_{x_2}=264.34$ GeV, $M_{H^+}=415.59$ GeV, $\sin\theta= 5.59 \times 10^{-4}$, $\mu^{''}_S=859.79$ GeV, $\lambda_{S12}=2.722$,  $\lambda_{S1}=2.5 \times 10^{-5}$, $\lambda_{S2}=1.163$, $\lambda_{S}=\lambda_2=\pi$, $\lambda_3= 0.5759$
(orange). The solid lines give the actual thermal history as described by the cBE system, while dotted the nBE one.}
\label{fig:Yyexample}
\end{center}
\end{figure}

In Fig.~\ref{fig:Yyexample} we show the thermal history of two example benchmark points for which kinetic decoupling happens early enough to have an impact on the final relic abundance. The left plot shows the evolution of the yield and how it differs between the cBE and the usual approach assuming kinetic equilibrium throughout (which we call nBE), while right plot highlights how the temperature evolution can differ from the one of the SM plasma. Both example points have strong semi-annihilation and kinetic decoupling happening at a time when the annihilations are still active, thus leading to a modification of the evolution of the DM number density. 

%%%%%%%%%%%%%%%%%%%%%%%%%%%%%%%%%%%%%%%%%%%%%%%%%%%%%%%%%%

\section{Phase transition and gravitational waves}
\label{sec:pt:gw}

The vacuum in which we live today corresponds to the minimum of the scalar potential located at the broken EW phase $(h,H,s)=(v,0,0)$. In the first moments of the Universe, however, the temperature was so large that thermal corrections influenced the $T=0$ effective potential significantly. As one goes back in time, the temperature raises and the increasing thermal effects push the low-temperature minimum (or true vacuum) upward: it becomes higher than the minimum at the origin and eventually disappears. Thus in the limit of  $T\rightarrow\infty$, the high-temperature, symmetric phase is usually $(h,H,s)=(0,0,0)$; the symmetry is restored. A more complicated possibility is symmetry non-restoration, where in this limit the minimum of the potential does not go to the origin~\cite{Mohapatra:1979qt, Mohapatra:1979vr}, or inverse symmetry breaking~\cite{Weinberg:1974hy}, where a symmetry, not broken at low temperature, is broken at high temperature.

In our case, the story 
 thus begins at the origin of the potential, in the high-temperature phase. Then as the temperature decreases, the potential develops a second minimum, say along the $h$ axis, where the potential is initially larger than at the origin. With time, the temperature continues to decrease until these two phases become degenerate: the value of the potential in both vacua is identical. The temperature at which it occurs is the critical temperature $T_c$. At some lower temperature, the field will eventually tunnel from this false vacuum to a deeper stable minimum (or true vacuum), here along the $h$ axis. This tunneling occurs through the nucleation of bubbles of true vacuum, which expand and collide with each others, filling the Universe little by little and thereby gradually converting the false vacuum into the true vacuum. The collision between these bubbles is responsible for the generation of a stochastic gravitational-wave background. The temperature at which one bubble nucleates per Hubble volume is called the nucleation temperature $T_n$. For a fast first-order phase transition without large reheating, this can be considered as the temperature at which gravitational waves are produced~\cite{Caprini:2015zlo}.\footnote{See. ref. \cite{Guo:2021qcq} for a discussion of subtleties.} Finally, the value of the Higgs VEV smoothly evolves towards its tree-level value $v$ as the temperature approaches zero.

A couple of important parameters capture the most information about a first-order phase transition. The strength of a first-order phase transition is characterized by the vacuum energy $\Delta \epsilon$ released during the phase transition. To obtain a dimensionless parameter $\alpha$, one normalizes this vacuum energy to the energy density of the plasma outside the bubbles $\rho_{\text{rad}}$~\cite{Espinosa:2010hh}:
\begin{equation}
\label{eq:alpha}
\alpha \equiv\frac{\Delta\epsilon}{\rho_{\text{rad}}}\Big|_{T=T_*},   \quad \Delta \epsilon \equiv \epsilon \big |_{\text{false vacuum}} - \epsilon \big |_{\text{true vacuum}}
\end{equation}
with
$\epsilon = V_{\rm eff} - \frac{T}{4}\frac{\partial V_{\rm eff}}{\partial T}$, $T_*$ the temperature at which GWs are generated and $\rho_{\text{rad}} = \frac{\pi^2}{30} g_* T^4$, where $g_*$ is the effective number of relativistic degrees of freedom at the temperature $T$. Note that a first-order phase transition is characterized as a strong one if the ratio of the VEV to the temperature, evaluated at $T_*$, satisfies
\begin{equation}
    \frac{v_*}{T_*}\geq 1.
\end{equation}
If this ratio is much smaller than unity the nature of the PT cannot be distinguished without lattice calculations, as also evidenced by the numerically very small value of the would-be GW signal in this case \cite{Kajantie:1996mn,Niemi:2020hto,Biekotter:2022kgf}.
The inverse time duration of the PT, $\beta$, is usually normalized by the Hubble parameter $H_*\equiv H(T_*)$~\cite{Grojean:2006bp}:
\begin{equation}
    \label{eq:beta}
    \frac{\beta}{H_*} = T\frac{d(S/T)}{dT}\Big |_{T=T_*},
\end{equation}
where $S\equiv S(h,H,s,T)$ is the 3-dimensional Euclidean action computed for an $O(3)$-symmetric critical bubble. A large $\beta/H_*$ means that we can safely neglect the Hubble expansion of the Universe during the phase-transition process.

A single bubble cannot generate gravitational waves, because its quadrupole moment vanishes due of its spherical symmetry. The key ingredient for a stochastic GW background  to be generated is the collisions between these bubbles of true vacuum, as mentioned above. Indeed, this will break the spherical symmetry of each of the collided scalar-field bubbles~\cite{Kamionkowski:1993fg}. For bubbles evolving in a thermal plasma, additional contributions 
 to the GW signal come from the metric perturbation induced by the plasma: sound waves~\cite{Hogan:1986qda, Hindmarsh:2013xza} and magnetohydrodynamic (MHD) turbulence~\cite{Caprini:2009yp}. Sub- and supersonic bubbles, respectively, create compression waves beyond the bubble wall and rarefaction waves behind it. When acoustic waves from different bubbles overlap, the induced shear stress of the metric generate gravitational waves. Finally, bubble collision is responsible for the turbulent motion of the fully ionized plasma. These MHD turbulences will generate gravitational waves. The power spectrum $h^2\Omega_{\text{GW}}$ of the stochastic GW background from a cosmic phase transitions thus  arises from three contributions~\cite{Caprini:2015zlo}: $h^2\Omega_{\text{GW}}\simeq h^2\Omega_{\text{col}} + h^2\Omega_{\text{sw}} + h^2\Omega_{\text{turb}}$.\footnote{In situations of strong supercooling, the GW power spectrum from sound-waves becomes similar to the contribution from bubble collision~\cite{Lewicki:2022pdb}.}
Note that the treatment of cosmic phase transitions suffers from uncertainties due to gauge or renormalisation-scale dependence for instance, that could potentially lower the resulting GW signal stength~\cite{Croon:2020cgk, Athron:2022jyi}.

The scalar-field contribution redshifted to today is given in the envelope approximation by~\cite{Huber:2008hg}\footnote{Recent works on contribution from bubble collisions can be found in~\cite{Lewicki:2019gmv, Lewicki:2020jiv, Lewicki:2020azd, Lewicki:2022pdb}.} 
\begin{equation}
 h^2\Omega_{\text{col}}(f) = h^2\Omega_{\text{col}}^{\text{peak}} S_\text{col}(f),
\end{equation}
with
\begin{equation}
\begin{split}
  h^2\Omega_{\text{col}}^\text{peak} &= 1.67\times 10^{-5}\left(\frac{H_*}{\beta}\right)^2\left(\frac{\kappa_\text{col}\alpha}{1+\alpha}\right)^2\left(\frac{100}{g_*}\right)^{1/3}\left(\frac{0.11 v_w^3}{0.42+v_w^2}\right),  
  \\
   \quad S_\text{col} &= \frac{3.8\left(f/f_\text{col}\right)^{2.8}}{1+2.8\left(f/f_\text{col}\right)^{3.8}},
\end{split}
\end{equation}
where $\kappa_\text{col}$ is the efficiency factor for the conversion of the vacuum energy into the gradient energy of the scalar field (energy stored in the shell of the scalar-field bubbles), $v_w$ is the bubble-wall speed in the rest frame of the plasma far away from the bubble~\cite{Caprini:2015zlo} and $f_\text{col}$ is the peak frequency, thus the frequency at $h^2\Omega_{\text{col}}^\text{peak}$.

The sound-wave contribution is given by~\cite{Hindmarsh:2017gnf, Caprini:2019egz, Schmitz:2020rag}

\begin{equation}
 h^2\Omega_{\text{sw}}(f) = h^2\Omega_{\text{sw}}^{\text{peak}} S_\text{sw}(f),
\end{equation}
with\footnote{The derivation of this ready-to-use formula is given in Appendix~\ref{sec:sw_contribution}.}
\begin{equation}
\begin{split}
h^2\Omega_{\text{sw}}^\text{peak} &= 1.23\times 10^{-6}\left(\frac{H_*}{\beta}\right)\left(\frac{\kappa_\text{sw}\alpha}{1+\alpha}\right)^2\left(\frac{100}{g_*}\right)^{1/3}v_w\Upsilon, 
\\
S_\text{sw} &= \left(\frac{f}{f_\text{sw}}\right)^3\left(\frac{7}{4+3\left(f/f_\text{sw}\right)^{2}}\right)^{7/2},
\end{split}
\end{equation}
where $\kappa_\text{sw}$ is the efficiency factor for the conversion of the vacuum energy into the bulk motion of the plasma and $f_\text{sw}$ is the sound-wave peak frequency.
The suppression factor for a radiation-dominated Universe is defined as~\cite{Guo:2020grp, Ellis:2019oqb} 
\begin{equation}
\Upsilon=1-\frac{1}{\sqrt{2\tau_\text{sw}H + 1}},
\end{equation}
where $\tau_\text{sw}$ is the time after which sound waves do not source the GW production and where $\Upsilon \simeq \tau_\text{sw}H$ when $\tau_\text{sw}H \ll 1$, thus when this time is much smaller than a Hubble time.
%%%%%%%%%%%%%%%%%%%%%%%%%%%%%%%%%%%%%%%%%%%%%%%%%%%%%%%%%%

Finally, the MHD-turbulence contribution is given by~\cite{Caprini:2015zlo, Schmitz:2020syl}
\begin{equation}
 h^2\Omega_{\text{turb}}(f) = h^2\Omega_{\text{turb}}^{\text{peak}} S_\text{turb}(f),
\end{equation}
with
\begin{align}
h^2\Omega_{\text{turb}}^\text{peak} &= 3.35\times 10^{-4}\left(\frac{H_*}{\beta}\right)\left(\frac{\kappa_\text{turb}\alpha}{1+\alpha}\right)^{3/2}\left(\frac{100}{g_*}\right)^{1/3}v_w\left[\frac{1}{2^{11/3}\left(1+8\pi f_\text{turb}/h_*\right)}\right], \\
S_\text{turb} &= \frac{\left(f/f_\text{turb}\right)^3}{\left[1+\left(f/f_\text{turb}\right)\right]^{11/3}} \left[\frac{2^{11/3}\left(1+8\pi f_\text{turb}/h_*\right)}{1+8\pi f/h_*}\right],
\end{align}
where $\kappa_\text{turb}$ is the efficiency factor for the conversion of the vacuum energy into the MHD turbulences, $f_\text{turb}$ is the MHD-turbulence peak frequency and $h_*$ is the today-redshifted value of the Hubble rate at $T_*$.

%%%%%%%%%%%%%%%%%%%%%%%%%%%%%%%%%%%%%%%%%%%%%%%%%%%%%%%%%%
\section{Numerical scan}
\label{sec:scan}

We used Markov Chain Monte Carlo (MCMC) with the Metropolis-Hastings algorithm in order to generate points, starting with a number of points with the DM relic density within the Planck $3 \sigma$ range. We then imposed other constraints: perturbativity, perturbative unitarity, bounded-from-below conditions for the scalar potential, globality of the EW vacuum, bounds on electroweak precision parameters $S$, $T$ and $U$, Higgs invisible width and Higgs to diphoton $h \to \gamma \gamma$ decay.

We performed three scans: 
\begin{enumerate}
\item Points with non-zero quartic $\lambda_{S12}$ coupling (and $\mu''_{S} = \lambda_{S1} = \lambda_{S2} = \lambda_{3} = 0$)
\item Points with non-zero cubic $\mu''_{S}$ coupling (and $\lambda_{S12} = \lambda_{S1} = \lambda_{S2} = \lambda_{3} = 0$)
\item General points with non-zero $\mu''_S$ and $\lambda_{S12}$ and without conditions on the quartic portals $\lambda_{S1}$, $\lambda_{S2}$, $\lambda_{3}$.
\end{enumerate}
In the first two cases, all the biquadratic portal couplings were set to zero, i.e. $\lambda_{S1} = \lambda_{S2} = \lambda_{3} = 0$, except for $\lambda_{4}$, because this would have entailed enforcing a special relation between $M_{x_{2}}$ and $M_{H^{\pm}}$. Since we need a non-zero mixing angle $\theta$ between $x_{1}$ and $x_{2}$, the $\mu_{SH}$ mixing term was non-zero in all cases. The dark sector scalar self-couplings were set to $\lambda_{S} = \lambda_{2} = \pi$ in order to allow for larger values of $\lambda_{S12}$ and $\mu''_{S}$ which are otherwise constrained by vacuum stability.

New points were generated from a multivariate normal distribution. The initial covariance matrix was calculated from a set of a few tens of points with DM relic density within the Planck $3\sigma$ bounds. The initial values for $M_{x_{2}}$ and $M_{H^{\pm}}$ were chosen in an interval around $2 M_{x_{1}}$ in order to enhance semi-annihilation processes. The covariance matrix was updated with each added point, while its norm was kept constant to ensure a reasonable acceptance ratio of points. Dimensionful quantities $M_{x_{1}}$, $M_{x_{2}}$, $M_{H^{\pm}}$ and $\mu''_{S}$ were generated logarithmically and only points with $M_{x_{1}} < M_{x_{2}}, M_{H^{\pm}}$ kept. Similarly to the calculation of the initial distribution, a few tens of points with DM relic density within the Planck $3\sigma$ bounds and varied couplings were used to start the MCMC chains from a few hundred up to thousand points in length. Since the desired probability distributions of our MCMCs were rather arbitrary, we do not consider the density of points in our plots to have a specific statistical meaning.

The desired distribution for each case consists of a product of Gaussians whose central values and standard deviations are given in Table~\ref{tab:mcmc}. Extra suppression factors of an arbitrary small value $10^{-12}$ were used to make the Gaussians essentially one-sided when the conditions $\sin \theta > 0$ and $0 \leq \lambda_{S12} < 4$ (the maximal value from vacuum stability bounds for $\lambda_{S} = \lambda_{2} = \pi$ and portal couplings $\lambda_{S1} = \lambda_{S2} = \lambda_{3} = 0$), and $1 \leq \mu''_{S}/\text{GeV} < 2 \sqrt{\lambda_{S}} M_{x_{1}}/\text{GeV}$ did not hold true. In the last condition, used in the second and third scans, we limited the cubic coupling by $\abs{\mu''_{S}} < 2 \sqrt{\lambda_{S}} M_{x_{1}}$, the bound from absolute vacuum stability in the $\mathbb{Z}_{3}$ singlet limit. Notice that unlike in the singlet-only limit, the cubic $\mu''_{S}$ coupling can yield semi-annihilation together with the ever-present mixing term $\mu_{SH}$. 

\begin{table}[tb]
\caption{The central values and standard deviations for the desired distribution of the MCMC.}
\begin{center}
\begin{tabular}{ccc}
  Quantity & Central value & Standard deviation \\
  \hline
  $\Omega_{c} h^{2}$ & $0.120$ & $0.001$ \\
  $M_{x_{1}}/\text{GeV}$ & $333.33$ & 500 \\
  $M_{x_{2}}/\text{GeV}$ & $2 M_{x_{1}}/\text{GeV}$ & 250 \\
  $M_{H^{\pm}}/\text{GeV}$ & $2 M_{x_{1}}/\text{GeV}$ & 250 \\
  $\sin \theta$ & $0$ & $0.125$ \\
  $\abs{M_{x_{2}} - M_{H^{\pm}}}/\text{GeV}$ & 0 & 360 \\
  $\mu''_{S}/\text{GeV}$ & $2 \sqrt{\lambda_{S}} M_{x_{1}}/\text{GeV}$ & $\frac{1}{3} \times 2 \sqrt{\lambda_{S}} M_{x_{1}}/\text{GeV}$
\end{tabular}
\end{center}
\label{tab:mcmc}
\end{table}%

%%%%%%%%%%%%%%%%%%%%%%%%%%%%%%%%%%%%%%%%%%%%%%%%%%%%%%%%%%
\section{Signals}
\label{sec:signals}

We now discuss the results of the scan, in particular for the direct-detection and gravitational-wave signals.

%%%%%%%%%%%%%%%%%%%%%%%%%%%%%%%%%%%%%%%%%%%%%%%%%%%%%%%%%%
\subsection{Direct detection}

\begin{figure}[t]
\begin{center}
  \includegraphics[width=0.5\linewidth]{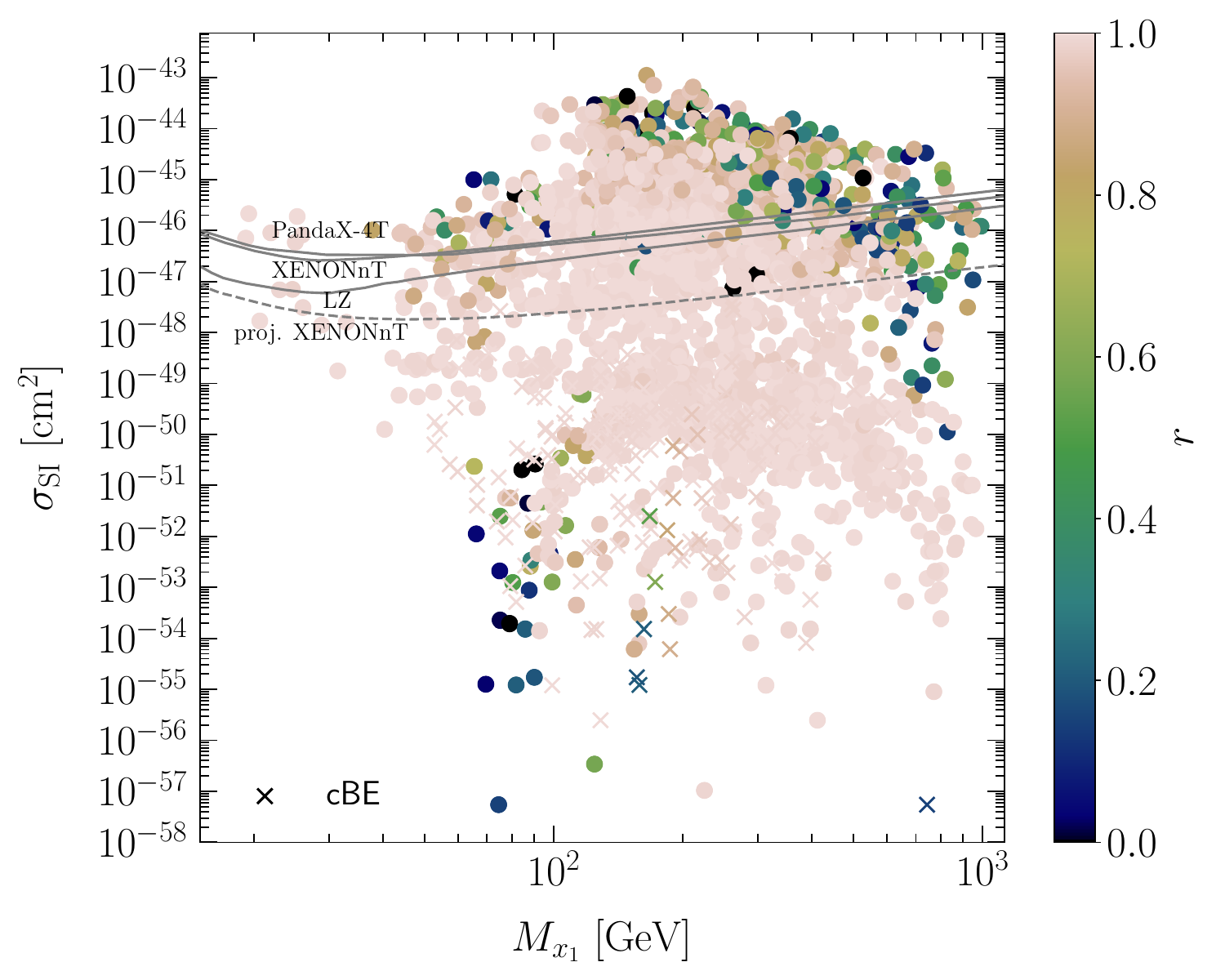}
  \\
  \includegraphics[width=0.5\linewidth]{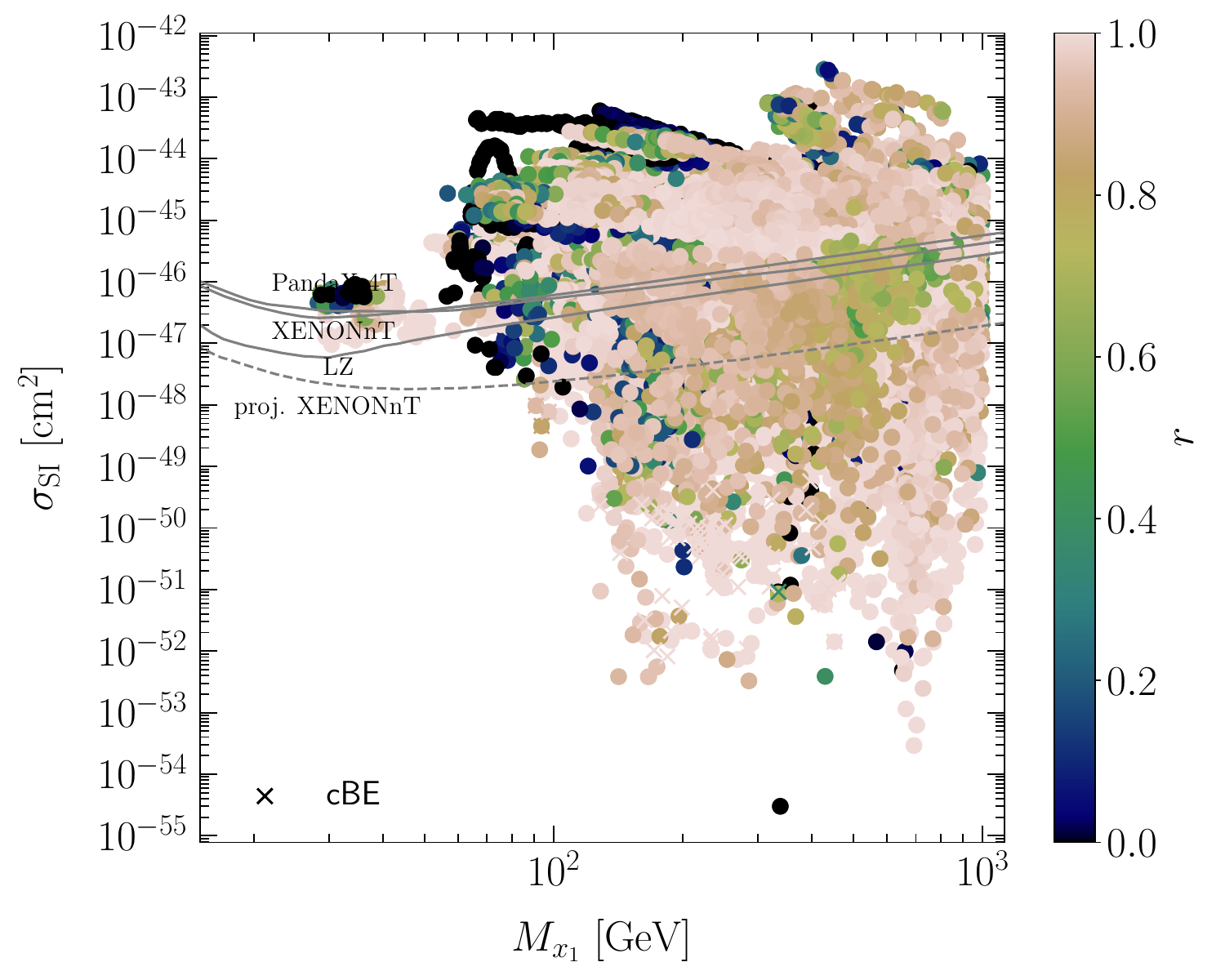}
  \\
  \includegraphics[width=0.5\linewidth]{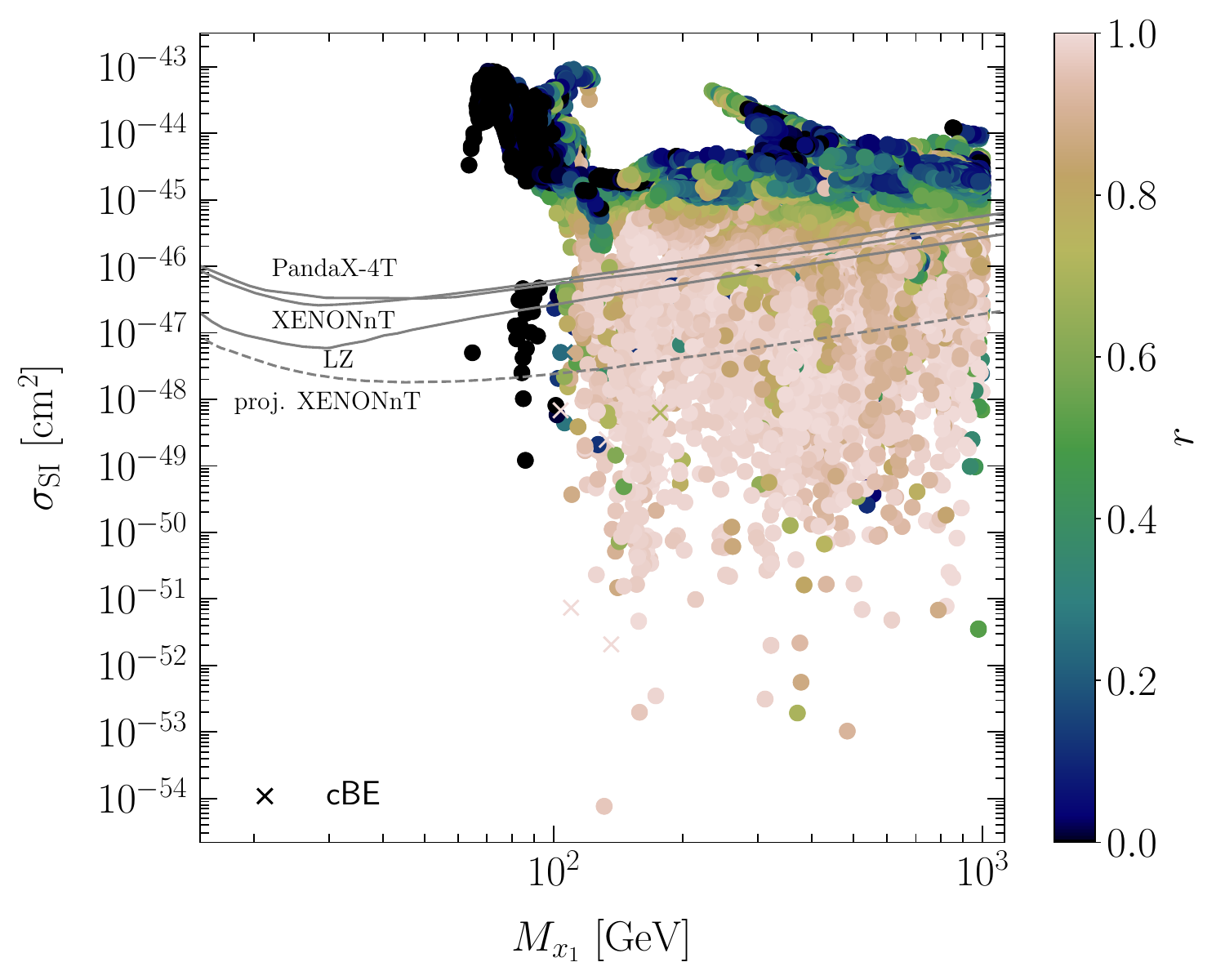}
\caption{Spin-independent direct detection cross section vs. dark matter mass. Top panel: points dominated by $\lambda_{S12}$; middle panel: points dominated by $\mu''_{S}$ and $\mu_{SH}$; bottom panel: more general points. Also shown are the sensitivity curves of the XENON1T (2018) \cite{XENON:2018voc}, PandaX-4T (2021) \cite{PandaX-4T:2021bab}, LZ (2022) \cite{LZ:2022ufs} in grey and the projected sensitivity curve for XENONnT \cite{XENON:2020kmp} in dashed grey.
}
\label{fig:direct:detection}
\end{center}
\end{figure}

In Fig.~\ref{fig:direct:detection} we show the predictions for the spin-independent direct detection cross section for the three cases. The top panel shows points with non-zero $\lambda_{S12}$, the middle panel points dominated by $\mu''_{S}$ and the bottom panel shows the points from the more general scan. Also shown are the sensitivity curves of the XENON1T (2018) \cite{XENON:2018voc}, PandaX-4T (2021) \cite{PandaX-4T:2021bab}, LZ (2022) \cite{LZ:2022ufs} in grey and the projected sensitivity curve for XENONnT \cite{XENON:2020kmp} in dashed grey. The color code shows the semi-annihilation fraction $r$ defined by
\begin{equation}
  r = \frac{1}{2} \frac{v \sigma_{x x \to x^{*} X}}{v \sigma_{x x^{*} \to X X} + \frac{1}{2} \sigma_{x x \to x^{*} X}},
\end{equation}
where $x$ denotes $x_{1}$, $x_{2}$, $H^{\pm}$ and $X$ stands for any SM particles. With dominant semi-annihilation $r$ approaches unity. The points with early kinetic decoupling, evaluated with the cBE method detailed in Section~\ref{sec:relic}, are marked with crosses.

%%%%%%%%%%%%%%%%%%%%%%%%%%%%%%%%%%%%%%%%%%%%%%%%%%%%%%%%%%
%\subsection{Indirect Detection}

%%%%%%%%%%%%%%%%%%%%%%%%%%%%%%%%%%%%%%%%%%%%%%%%%%%%%%%%%%
\subsection{Gravitational waves}
\label{sec:GW}

\begin{figure}[tb]
\begin{center}
  \includegraphics[width=0.49\linewidth]{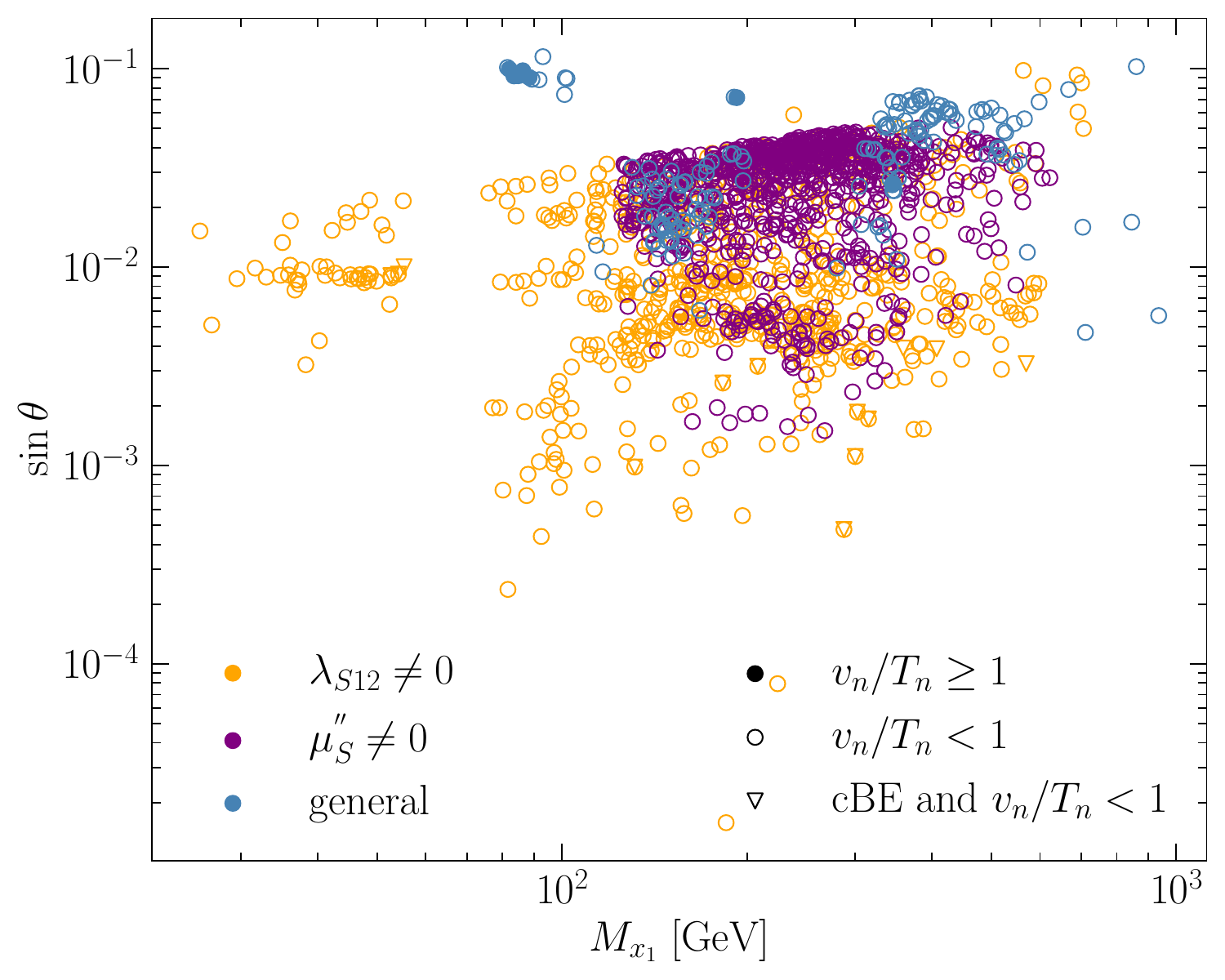}
  \includegraphics[width=0.49\linewidth]{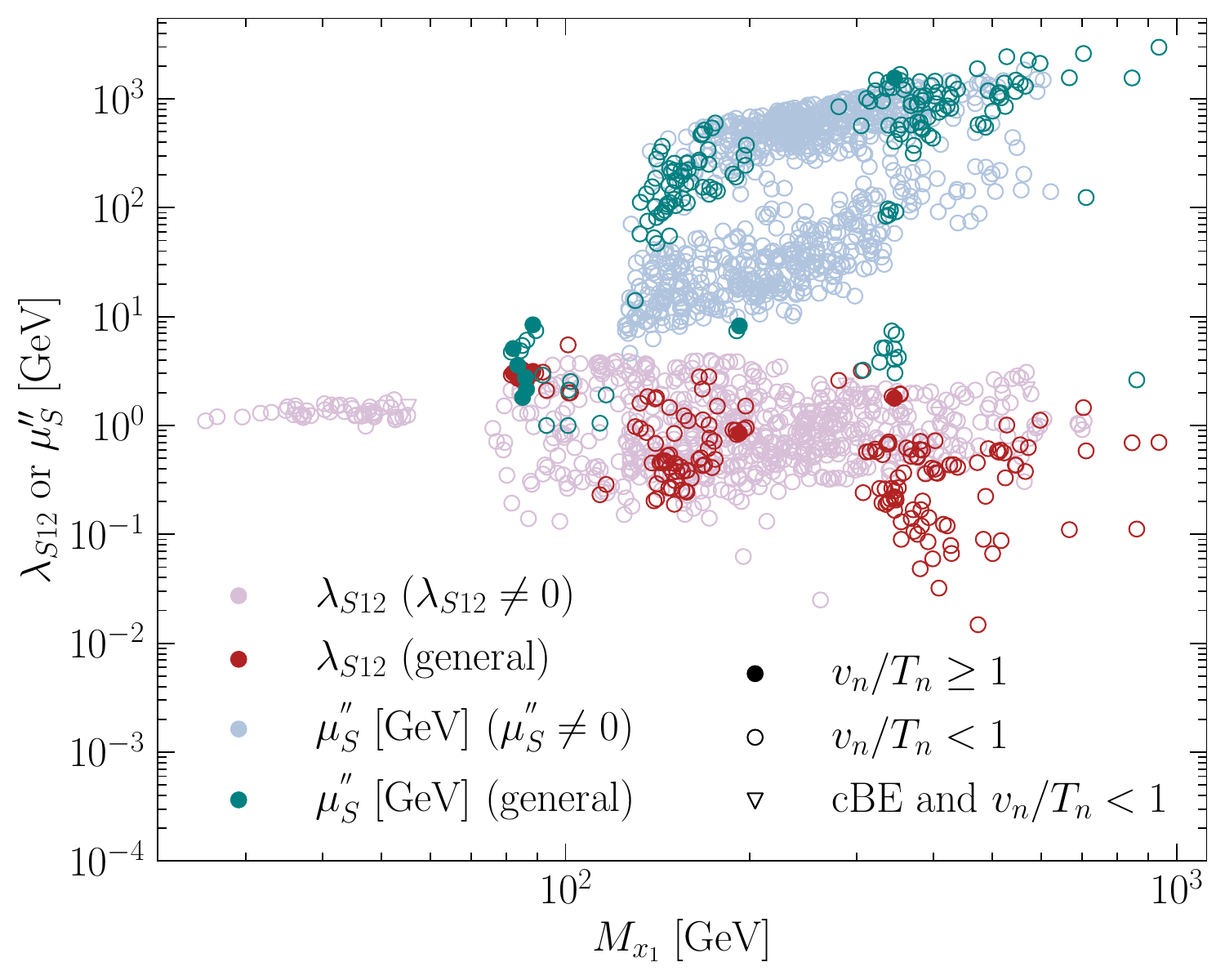}
\caption{Projection of the parameter space, showing points which satisfy all the theoretical/experimental constraints. %\AH{Not sure we want to highlight these points with separate markers and emphsize in the catpion:}
cBE points that lead to FOPT are only found for $\lambda_{S12}\neq 0$ scan. Strong FOPT are only found for general scan. Left panel: Sine of the mixing angle $\sin\theta$ vs. DM mass $M_{x_1}$. Right panel: $\lambda_{S12}$ vs. $M_{x_1}$ for both the  $\lambda_{S12}\neq 0$ and general scan and $\mu_S''$ vs. $M_{x_1}$ for both the  $\mu''_S\neq 0$ and general scan.}
\label{fig:mx1_sinTheta}
\end{center}
\end{figure}

\begin{figure}[p]
\begin{center}
  \includegraphics[width=0.5\linewidth]{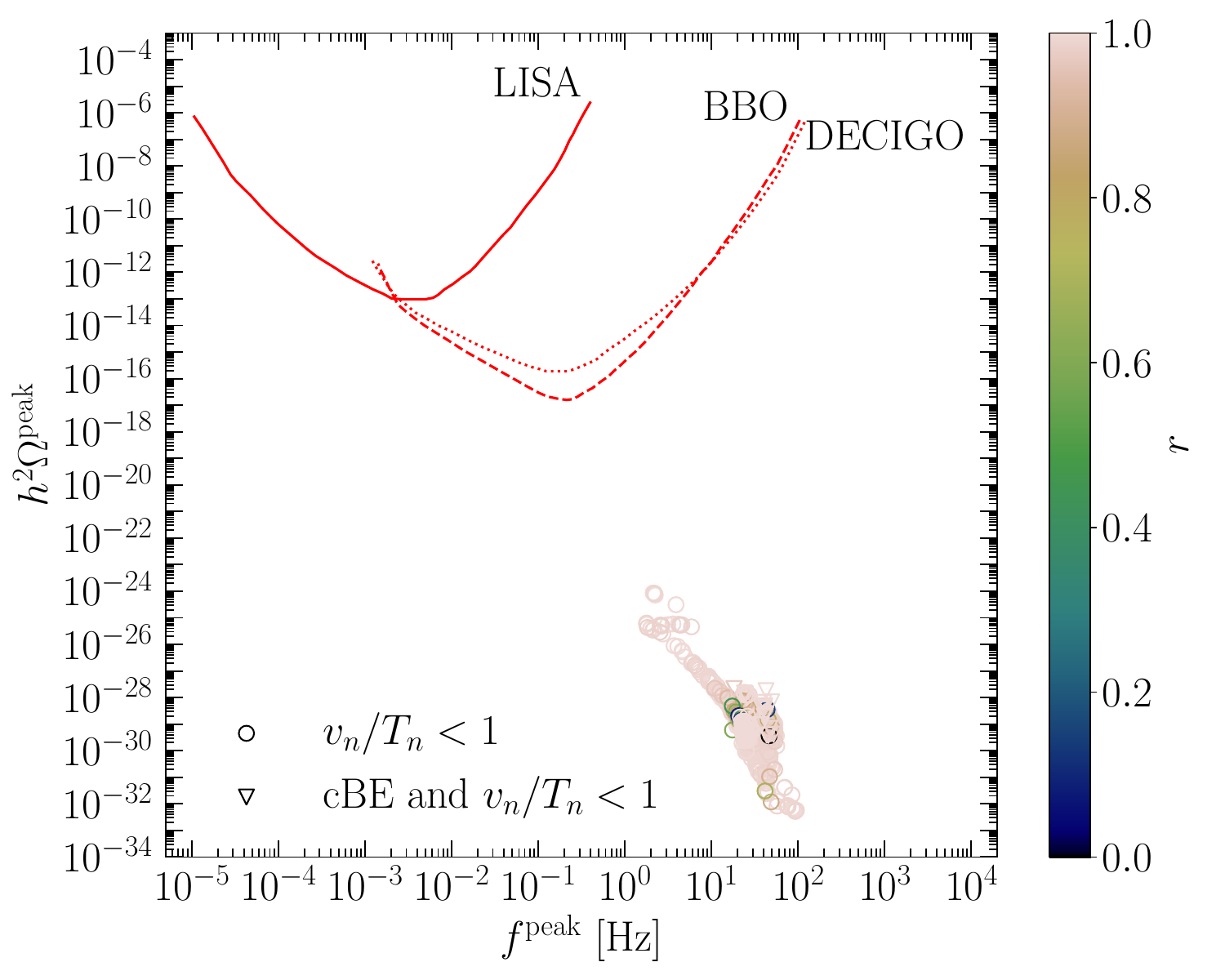}\\
    \includegraphics[width=0.5\linewidth]{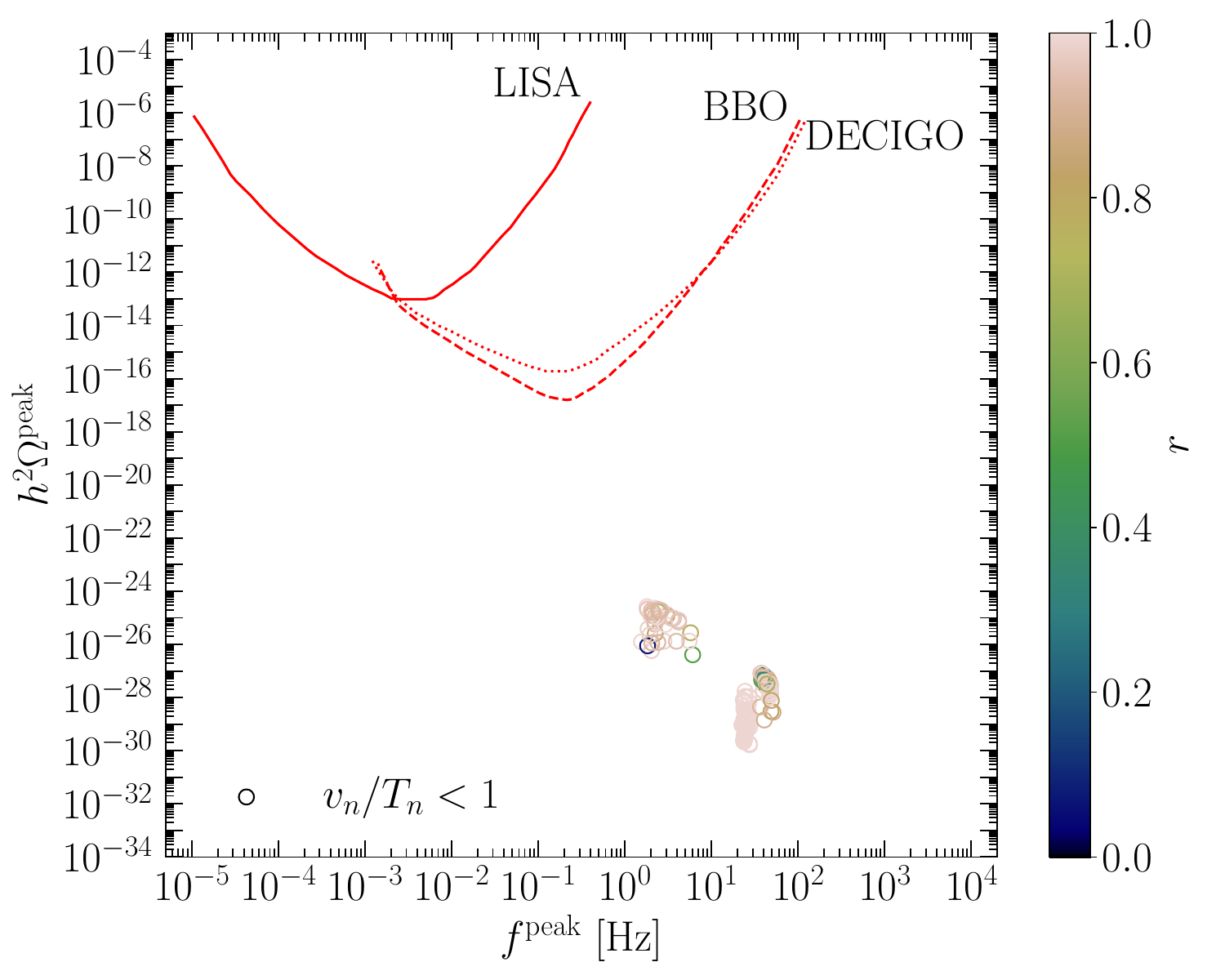}\\ 
     \includegraphics[width=0.5\linewidth]{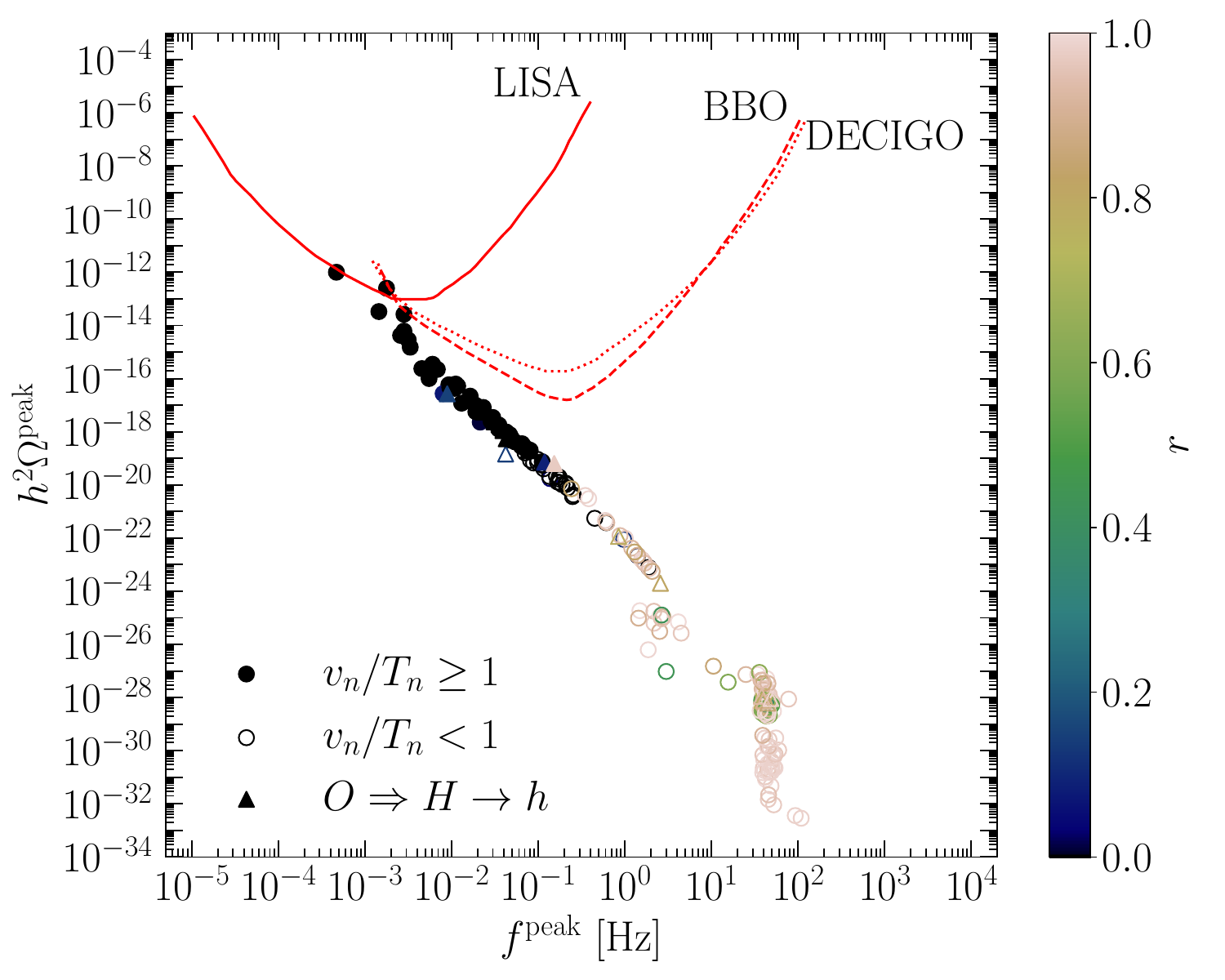}
\caption{Peak amplitude of the GW signal as a function of peak frequency. The top panel shows points with non-zero $\lambda_{S12}$, the middle panel points dominated by $\mu''_{S}$ and the bottom panel shows the points from the more general scan. Also shown are the PLI sensitivity curves of future detectors LISA, BBO and DECIGO.}
\label{fig:gw_signal}
\end{center}
\end{figure}

For each of the three scans mentioned in Section~\ref{sec:scan}, we study how the Universe ends in the EW vacuum $(v, 0, 0)$ at zero temperature. If at least one phase transition, to reach that state, is of the first order, then we subsequently compute the power spectrum of the stochastic GW background $h^2\Omega_{\text{GW}}$, necessarily associated to these cosmic FOPTs. In addition to the previous constraints, we also take into account the constraint from direct-detection experiments: only points not excluded by LZ are displayed in Fig.~\ref{fig:mx1_sinTheta} and Fig.~\ref{fig:gw_signal}. Note also that we consider a relativistic bubble-wall speed, $v_w=1$.

In the left panel of Fig.~\ref{fig:mx1_sinTheta}, we show, for the three scans, the region of the parameter space, projected onto the plane made by the mixing angle $\sin \theta$ and the DM mass $M_{x_1}$, that gives rise to first-order phase transitions. These are essentially obtained for large mixing and mild to large DM mass. We see that only points from the general scan lead to strong FOPTs. The right panel shows the values of either $\lambda_{S12}$ or $\mu_S''$, as a function of $M_{x_1}$, responsible for FOPTs for scans where they are non-zero. In both panels, triangles are points for which the cBE method has been used to compute the DM relic density, as the scattering rate is not sufficient to consider kinetic equilibrium for DM during the freeze-out process (see Section~\ref{sec:relic}).

The signal from the stochastic gravitational-wave background is shown in Fig.~\ref{fig:gw_signal}, with the power-law integrated (PLI) sensitivity curves of LISA, DECIGO and BBO, obtained for an observation time of 4 years and a threshold value for the signal-to-noise ratio to be 10~\cite{Thrane:2013oya, Azatov:2019png}. % in the left column of Fig.~\ref{fig:gw_signal}.
Similarly to Fig.~\ref{fig:direct:detection}, we show the semi-annihilation fraction with a color code. The upper panel of Fig.~\ref{fig:gw_signal} shows the GW signal for the scan with non-zero $\lambda_{S12}$. This signal results from single-step FOPTs. We can see that none of the points leads to strong FOPTs and that for some of them (triangles), the assumption of kinetic equilibrium does not hold and the full cBE system should be solved. The middle panel shows the GW signal for the scan with non-zero $\mu''_{S}$. Again, we only find single-step weak FOPTs. A cubic term helps to generate a barrier in the potential and thus enhance the strength of the phase transition. However, we only find a single-step FOPT from the origin towards the EW minimum, thereby along the Higgs direction $h$. This is why a non-zero $\mu''_{S}$ is not helpful here to obtain a stronger PT, as it only impacts the $s$ direction. Finally, the lower panel of Fig.~\ref{fig:gw_signal} shows the GW signal for the general scan, where strong FOPTs are obtained and where one of the points clearly lies above the LISA PLI sensitivity curve, thus yielding a detectable GW signal. Note that we also found two-step phase transitions. For each of them, indicated by a triangle, the first step $(0,0,0)\rightarrow (0,v_H,0)$ is a second-order PT and is followed by a first-order PT $(0,v_H^{'},0)\rightarrow (v_h,0,0)$. Note also that from Fig.~\ref{fig:gw_signal}, it seems that the lower the semi-annihilation fraction, the stronger the GW signal. However, if we do not take experimental constraints into account, the correlation between the strength of the GW signal and the semi-annihilation fraction is lost.

\begin{figure}[h!]
\begin{center}
  \includegraphics[width=0.52\linewidth]{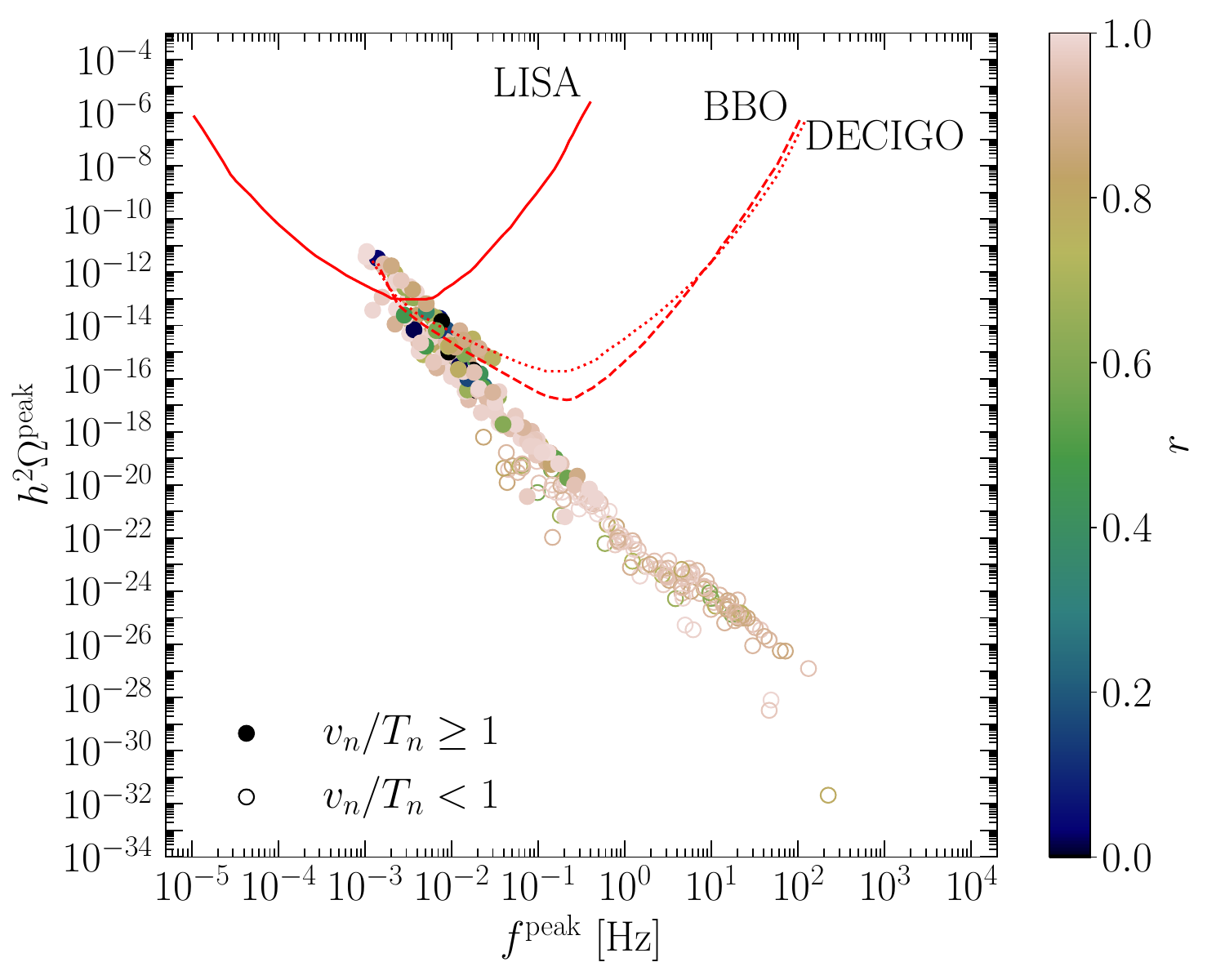}
    \includegraphics[width=0.47\linewidth]{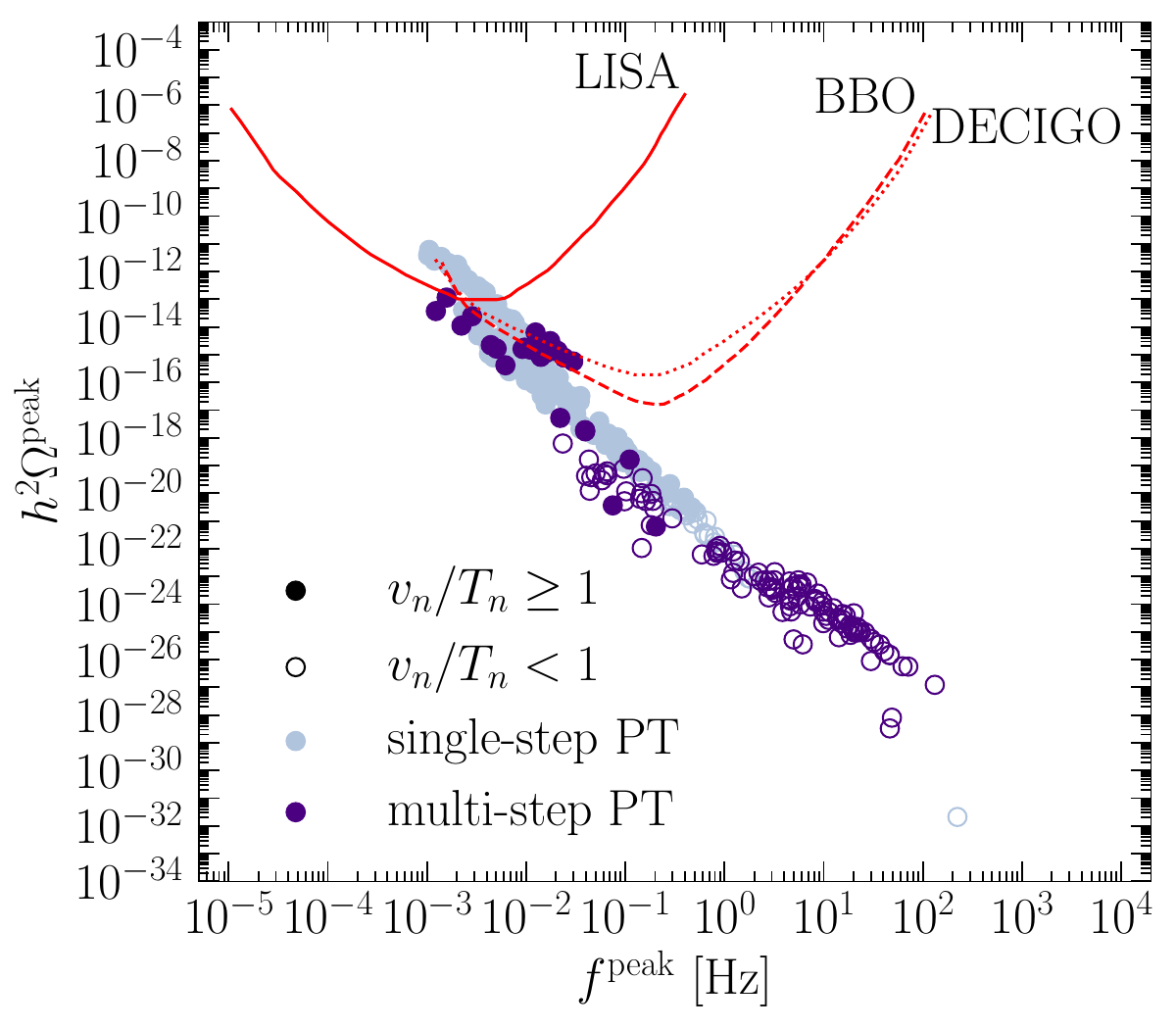}
\caption{Peak amplitude of the GW signal as a function of peak frequency with points from the scan in Eq.~(\ref{eq:fully_general_scan}). Also shown are the PLI sensitivity curves of future detectors LISA, BBO and DECIGO. The left panel highlights the semi-annihilation rate, while the right panel distinguishes single-step from multi-step PTs.}
\label{fig:gw_signal_general}
\end{center}
\end{figure}

Finally, let us stress that in principle, in this model, if we relax the constraint on relic density and allow for underabundant DM, we can more easily find regions of the parameter space allowing for different kind of multi-step phase transitions in the three-dimensional field space $(h,H,s)$ and/or leading to a strong enough GW signal to be detected by LISA, DECIGO or BBO, as can be seen in Fig.~\ref{fig:gw_signal_general}. The points in this figure comply with all the theoretical/experimental constraints, except that they only lead to underabundant DM, and result from a random scan in the following range:

\begin{align}
\label{eq:fully_general_scan}
 M_{x_1} &\in [10, 1000]\text{~GeV}, & M_{x_2}, M_{H^\pm} &\in [M_{x_1}, 1000\text{~GeV}],
 &
 \sin\theta &\in [0, \sqrt{2}/2],
\notag
\\
 \mu''_S &\in [0, 2000]\text{~GeV}, 
&
\lambda_3, \lambda_{S1},\lambda_{S2},\lambda_{S12} &\in [-4\pi, 4\pi], 
&
\lambda_{2},\lambda_{S} &\in [-\pi, 4\pi].
\end{align}

%%%%%%%%%%%%%%%%%%%%%%%%%%%%%%%%%%%%%%%%%%%%%%%%%%%%%%%%%%
\section{Conclusions}
\label{sec:conclusions}

We have explored the cosmology of the $\mathbb{Z}_{3}$ symmetric dark matter model with an inert doublet and a complex singlet with the emphasis on detectability of the gravitational-wave signal and dark-matter phenomenology. In order for the model to be in agreement with direct-detection limits the dark matter candidate needs to be a singlet-like state arising from the mixing of the singlet and the neutral part of the inert doublet. This setup opens an interesting possibility of important or even dominant semi-annihilation processes that modifies the thermal DM production through an early kinetic decoupling: if the relic density is determined via semi-annihilation, the dark matter-Higgs boson couplings that would keep DM in kinetic equilibrium via elastic collisions are suppressed.

We performed MCMC scans to not only produce general coupling configurations available to the model, but also to look specifically at large semi-annihilation (through a large cubic coupling $\mu''_{S}$ or the quartic semi-annihilation coupling $\lambda_{S12}$ coupling). We find that for the points with dominant semi-annihilation, strong suppression of the direct-detection signal comes also with weak FOPTs whose GW signals would lie orders of magnitude below detection projections in a foreseeable future. On the other hand, a general scan of the model reveals points near the detection limits of LISA, DECIGO and BBO.

We then also computed the GW signal for uniformly distributed random general scan  allowing for underabundant singlet-like $x_1$ with relic density below the Planck constraint. (There could be either an additional non-thermal production mechanism or another DM component.) In that case we found a sizable portion of the parameter space leading to a single-step or a multi-step phase transition producing strong enough GW signal to be potentially detected by LISA, DECIGO or BBO.

Therefore, we conclude that in the $\mathbb{Z}_{3}$ symmetric model with an inert doublet and a complex singlet there exists a parameter range that explaining whole dark matter \textit{and} providing potentially observable GW peak amplitude, though it is rather small and challenging to be observed in the near future. Nevertheless, when extending the analysis to underabundant dark matter the detection prospects of the GW peak amplitude become significantly more promising.

%%%%%%%%%%%%%%%%%%%%%%%%%%%%%%%%%%%%%%%%%%%%%%%%%%%%%%%%%%
\appendix

%%%%%%%%%%%%%%%%%%%%%%%%%%%%%%%%%%%%%%%%%%%%%%%%%%%%%%%%%%

\section{Field-dependent scalar mass matrices}
\label{sec:mass_matrix}
The field-dependent scalar masses $m_i^2\equiv m_i^2(h,H,s)$, present in Eq.~(\ref{eq:Vcw}) and Eq.~(\ref{eq:VT}), correspond to the eigenvalues of the field-dependent mass matrices of the scalar fields $M^2_S$, of the pseudoscalar fields $M^2_P$ and of the charged fields $M^2_C$:
\begin{align}
M_{S}^2 &= {\footnotesize
\begin{pmatrix}
 \mu_{1}^2 + 3\lambda_1 h^2 + \frac{\lambda_3+\lambda_4}{2} H^2 + \frac{\lambda_{S1}}{2}s^2 &(\lambda_3+\lambda_4)hH + \frac{\lambda_{S12}s+\sqrt{2}\mu_{SH}}{4}s & \lambda_{S1}hs+\frac{\lambda_{S12}}{2}Hs+\frac{\sqrt{2}\mu_{SH}}{4}H\\
(\lambda_3+\lambda_4)hH + \frac{\lambda_{S12}s+\sqrt{2}\mu_{SH}}{4}s & \mu_{2}^2 + 3\lambda_2 H^2 + \frac{\lambda_3+\lambda_4}{2} h^2 + \frac{\lambda_{S2}}{2}s^2  & \lambda_{S2}Hs+\frac{\lambda_{S12}}{2}hs+\frac{\sqrt{2}\mu_{SH}}{4}h\\
 \lambda_{S1}hs+\frac{\lambda_{S12}}{2}Hs+\frac{\sqrt{2}\mu_{SH}}{4}H  & \lambda_{S2}Hs+\frac{\lambda_{S12}}{2}hs+\frac{\sqrt{2}\mu_{SH}}{4}h &\mu_{S}^2 + 3\lambda_S s^2+ \frac{3}{\sqrt{2}}\mu''_S s\\
 & & + \frac{\lambda_{S1}h^2+\lambda_{S2}H^2+\lambda_{S12}hH}{2}
\end{pmatrix}},
\\
M_{P}^2 &= {\footnotesize
\begin{pmatrix}
 \mu_{1}^2 + \lambda_1 h^2 + \frac{\lambda_3+\lambda_4}{2} H^2 + \frac{\lambda_{S1}}{2}s^2 & \frac{\lambda_{S12}s+\sqrt{2}\mu_{SH}}{4}s & \frac{\lambda_{S12}}{2}Hs-\frac{\sqrt{2}\mu_{SH}}{4}H\\
 \frac{\lambda_{S12}s+\sqrt{2}\mu_{SH}}{4}s & \mu_{2}^2 + \lambda_2 H^2 + \frac{\lambda_3+\lambda_4}{2} h^2 + \frac{\lambda_{S2}}{2}s^2  &-\frac{\lambda_{S12}}{2}hs+\frac{\sqrt{2}\mu_{SH}}{4}h\\
 \frac{\lambda_{S12}}{2}Hs-\frac{\sqrt{2}\mu_{SH}}{4}H  & -\frac{\lambda_{S12}}{2}hs+\frac{\sqrt{2}\mu_{SH}}{4}h &\mu_{S}^2 + \lambda_S s^2- \frac{3}{\sqrt{2}}\mu''_S s\\
 & & + \frac{\lambda_{S1}h^2+\lambda_{S2}H^2-\lambda_{S12}hH}{2}
\end{pmatrix}},
\\
M_{C}^2 &= 
\begin{pmatrix}
 \mu_{1}^2 + \lambda_1 h^2 + \frac{\lambda_3}{2} H^2 + \frac{\lambda_{S1}}{2}s^2 & \frac{\lambda_4}{2} hH+\frac{\lambda_{S12}s+\sqrt{2}\mu_{SH}}{4}s \\
 \frac{\lambda_4}{2} hH+\frac{\lambda_{S12}s+\sqrt{2}\mu_{SH}}{4}s & \mu_{2}^2 + \lambda_2 H^2 + \frac{\lambda_3}{2} h^2 + \frac{\lambda_{S2}}{2}s^2
\end{pmatrix}.  
\end{align}

Evaluated in the EW vacuum $(h,H,s)=(v,0,0)$, these mass matrices become
\begin{align}
M_{S}^2 &= 
\begin{pmatrix}
 \mu_{1}^2 + 3\lambda_1 v^2 & 0 & 0\\
0 & \mu_{2}^2 + \frac{\lambda_3+\lambda_4}{2} v^2 & \frac{\sqrt{2}\mu_{SH}}{4}v\\
 0  & \frac{\sqrt{2}\mu_{SH}}{4}v &\mu_{S}^2 +  \frac{\lambda_{S1}v^2}{2}
\end{pmatrix}, 
\\
M_{P}^2 &= 
\begin{pmatrix}
 \mu_{1}^2 + \lambda_1 v^2 & 0 & 0\\
 0& \mu_{2}^2 +\frac{\lambda_3+\lambda_4}{2} v^2 &\frac{\sqrt{2}\mu_{SH}}{4}v\\
0  & \frac{\sqrt{2}\mu_{SH}}{4}v &\mu_{S}^2 +\frac{\lambda_{S1}v^2}{2}
\end{pmatrix},
\\
M_{C}^2 &= 
\begin{pmatrix}
 \mu_{1}^2 + \lambda_1 h^2 & 0 \\
0 & \mu_{2}^2 +  \frac{\lambda_3}{2} v^2 
\end{pmatrix}, 
\end{align}
thus putting in evidence the mixing between $H_2$ and $S$.

\section{Counter-terms}
\label{sec:counterterms}

We present expressions for counter-terms in the effective potential used to keep VEVs, masses and mixing at their tree-level values. Since the counter-terms $\delta\lambda_3$ and $\delta\lambda_4$ only appear as a sum in Eq.~(\ref{eq:Vct}), we can arbitrarily set $\delta\lambda_4$ to zero. The counter-terms, expressed via derivatives of the one-loop Coleman-Weinberg potential \eqref{eq:Vcw}, are given by
\begin{align}
  \delta \mu_{1}^{2} &= \frac{v\partial_{h}^{2} \Delta V  - 3\partial_{h} \Delta V}{4v},
  \\
    \delta \mu_{2}^{2} &= \frac{v \partial_{h}\partial_{H^0}^2  V - 2\partial_{H^0}^{2} \Delta V}{4},
  \\
  \delta \mu_{S}^{2} &= \frac{v \partial_{h}\partial_{s_R}^2  V - 2\partial_{s_{R}}^{2} \Delta V}{4},
  \\
  \delta \mu_{SH} &= -\partial_{H^{0}} \partial_{s_{R}} \Delta V,
  \\
  \delta \mu''_{S} &= -\frac{1}{6} \partial_{s_R}^{3} \Delta V,
  \\
  \delta \lambda_{1} &= \frac{\partial_{h} \Delta V  - v\partial_{h}^2 \Delta V}{8v^3},
  \\
  \delta \lambda_{3} &= -\frac{1}{4v} \partial_{h}\partial_{H^0}^2 \Delta V,
  \\
  \delta \lambda_{S1} &= -\frac{1}{4v} \partial_{h}\partial_{s_R}^2 \Delta V,
  \\
  \delta \lambda_{S12} &= - \frac{1}{2v}\partial_{H^0} \partial_{s_R}^2 \Delta V.
\end{align}

%%%%%%%%%%%%%%%%%%%%%%%%%%%%%%%%%%%%%%%%%%%%%%%%%%%%%%%%%%
\section{Sound-wave contribution $h^2\Omega_{\text{sw}}(f)$}
\label{sec:sw_contribution}

From~\cite{Hindmarsh:2017gnf, Caprini:2019egz, Schmitz:2020rag} one has\footnote{Note that in~\cite{Caprini:2019egz} the factor 3 is missing in Eq.~(\ref{eq:omega_sw}) and the factor $1/c_s$ is not considered in~\cite{Hindmarsh:2017gnf, Schmitz:2020rag}.}
\begin{equation}
\label{eq:omega_sw}
h^2\Omega_{\text{sw}}^\text{peak} = 3 h^2 \times 0.687 F_\text{gw,0}\Gamma^2\bar{U}_f^4 \frac{H_* R_*}{c_s} \tilde{\Omega}_\text{gw}\Upsilon, 
\end{equation}
where the mean adiabatic index $\Gamma=4/3$ for a relativistic fluid, and the enthalpy-weighted root-mean-square of the fluid velocity $\bar{U}_f$ is expressed as~\cite{Schmitz:2020rag}:
\begin{equation}
\bar{U}_f = \sqrt{\frac{K}{\Gamma}}, \quad K=\frac{\kappa_\text{sw}\alpha}{1+\alpha}.
\end{equation}

The mean bubble separation at percolation temperature\footnote{Here we consider that the nucleation and percolation temperature are very close to each other: $T_n\simeq T_p$.} $R_*$ is defined as
\begin{equation}
R_* = \frac{(8\pi)^{1/3}\text{max}\{c_s, v_w\}}{\beta}
\end{equation}
with $c_s=1/\sqrt{3}$ the speed of sound in the plasma~\cite{Espinosa:2010hh}.

The parameter $\tilde{\Omega}_\text{gw}$, which quantifies how efficiently the kinetic energy is converted into gravitational waves, is numerically found to be $\tilde{\Omega}_\text{gw}\simeq 1.2\times 10^{-2}$~\cite{Hindmarsh:2017gnf, Schmitz:2020rag}.

Finally, the factor $F_\text{gw,0}$ needed to obtain the today value $h^2\Omega_{\text{sw}}^\text{peak}$ is defined as~\cite{Caprini:2019egz, Schmitz:2020rag}
\begin{equation}
F_\text{gw,0} = \Omega_{\gamma,0}\left(\frac{g_{s0}}{g_{s*}}\right)^{4/3}\frac{g_*}{g_0},
\end{equation} 
where $\Omega_{\gamma,0}\equiv\rho_{\gamma,0}/\rho_{c,0}$ is the photon density today, $g_{s*}$ the effective number of entropic degrees of freedom, $g_{*}$ the effective number of relativistic degrees of freedom, while $g_{s0}$ and $g_{0}$ give their present values, respectively. The degrees of freedom $g_{s*}$ and  $g_{*}$ are equal at the EW scale, while $g_{s0}$ and $g_{0}$ are different since they are considered after the neutrino decoupling.

The following values for the aforementioned parameters allows us to compute the factor $F_\text{gw,0}$~\cite{Mulders:2019vhb, Gelmini:1996kv}:
\begin{align}
 &\rho_{\gamma,0} = 0.26~\frac{\text{eV}}{\text{cm}^3},\quad \rho_{c,0}=1.05\times 10^{-5}h^2~\text{GeV}\, \text{cm}^{-3},\quad h\equiv \frac{H_0}{100~\text{km~s}^{-1}\text{Mpc}^{-1}},\nonumber\\
 & g_{s0} \simeq 3.94\quad\text{and} \quad g_0 \simeq 3.38,
\end{align}
where $N_\text{eff}=3.044$~\cite{Akita:2022hlx} has been used in the computation of $g_{s0}$ and $g_0$. Note that $g_{s0} \simeq 3.91$ and $g_0=2$ are considered in Refs.~\cite{Hindmarsh:2017gnf, Caprini:2019egz}. Considering these values, we can thus express $F_\text{gw,0}$ as
\begin{equation}
 F_\text{gw,0} = \frac{9.77\times 10^{-6}}{h^2}\left(\frac{100}{g_*}\right)^{1/3}
\end{equation}

Putting all together, we finally obtain 
\begin{align}
 h^2\Omega_{\text{sw}}^\text{peak} &= 2.061h^2\frac{9.77\times 10^{-6}}{h^2}\left(\frac{100}{g_*}\right)^{1/3}\left( \frac{\kappa_\text{sw}\alpha}{1+\alpha}\right)^2 \frac{H_*(8\pi)^{1/3}\text{max}\{c_s, v_w\}}{\beta/\sqrt{3}}1.2\times 10^{-2}~\Upsilon \nonumber\\
&\simeq 1.23\times 10^{-6}\left(\frac{H_*}{\beta}\right)\left( \frac{\kappa_\text{sw}\alpha}{1+\alpha}\right)^2\left(\frac{100}{g_*}\right)^{1/3}\text{max}\{c_s, v_w\}~\Upsilon.
\end{align}
%%%%%%%%%%%%%%%%%%%%%%%%%%%%%%%%%%%%%%%%%%%%%%%%%%%
%\acknowledgments

\section*{Acknowledgements}

This work was supported by the Estonian Research Council grants PRG434, PRG803, RVTT3 and RVTT7, and by the CoE program TK202 ``Fundamental Universe''. The work of A.H. was supported by the National Science Centre, Poland, research grant No. 2021/42/E/ST2/00009. The work of M.L. is supported by the National Science Centre, Poland, research grant No. 2020/38/E/ST2/00243.

%\newpage

%%%%%%%%%%%%%%%%%%%%%%%%%%%%%%%%%%%%%%%%%%%%%%%%%%%
\bibliographystyle{JHEP}
\bibliography{ZNSIDSM.bib}

\end{document}